# Choice probabilities and correlations in closed-form route choice models: specifications and drawbacks

Fiore Tinessa*[1], Vittorio Marzano[1], Andrea Papola[1]

[1]*Università degli Studi di Napoli Federico II*
*Dipartimento di Ingegneria Civile, Edile e Ambientale (DICEA)*
*Via Claudio 21, 80125 Naples (Italy)*

* corresponding author: fiore.tinessa@unina.it

October 2021

**Abstract**

*This paper investigates the performance – in terms of choice probabilities and correlations – of existing and new specifications of closed-form route choice models with flexible correlation patterns, namely the Link Nested Logit (LNL), the Paired Combinatorial Logit (PCL) and the more recent Combination of Nested Logit (CoNL) models. Following a consolidated track in the literature, choice probabilities and correlations of the Multinomial Probit (MNP) model by (Daganzo & Sheffi, 1977) are taken as target. Laboratory experiments on small/medium-size networks are illustrated, also leveraging a procedure for practical calculation of correlations of any GEV models, proposed by (Marzano 2014). Results show that models with inherent limitations in the coverage of the domain of feasible correlations yield unsatisfactory performance, whilst the specifications of the CoNL proposed in the paper appear the best in fitting both MNP correlations and probabilities. Performance of the models are appreciably ameliorated by introducing lower bounds to the nesting parameters. Overall, the paper provides guidance for the practical application of tested models.*

## 1. Motivation and setup

Traffic assignment procedures represent the most common tool to turn origin-destination (o-d) demand flows into vehicles/pedestrians link flows in transport networks (see Cascetta 2009; Ortúzar and Willumsen 2011). A key component of traffic assignment procedures is represented by route choice models, aimed to split o-d flows amongst available routes for each o-d pair. Modelling route choice is indeed a well-established research track in transport engineering, with a consolidated state-of-the-art tackling a variety of relevant concerned research issues: comprehensive reviews are provided, amongst others, by Ramming (2002), Prato and Bekhor (2007), Prato (2009), Papola and Marzano (2013), Papola, Tinessa, and Marzano (2018), Simonelli, Tinessa, Buonocore, & Pagliara (2020).

The paradigm of discrete choice modelling has been applied to route choice by researchers and practitioners. A review of the concerned literature reveals that the most of the proposed discrete choice models for route choice is consistent with the following assumptions, useful also to focus the research framework of the paper.

*Definition of route as choice alternative.* A route is commonly defined as an ordered sequence of links connecting an origin-destination (o-d) pair, without any loops, leading to a finite route choice set and to standard traffic assignment procedures (STA), also implemented in commercial software.

In fact, a relatively more recent research track allows presence of loops in routes, leading to a choice set with infinite cardinality and to Markovian traffic assignment procedures (Bell 1995; Akamatsu 1996; Baillon and Cominetti 2007; Fosgerau, Frejinger, and Karlstrom 2013; Mai, Fosgerau, and Frejinger 2015; Mai 2016; Oyama and Hato 2019). This paper resorts to the loop-free definition, consistent with most of the papers enabling comparisons of performance amongst route choice models in terms of route choice probabilities and correlations, as recalled in Section 2.

*Decision rule*. Different paradigms have been applied to route choice modelling, i.e.: prospect theory (Kahneman & Tversky, 1979; Huang, Burris, & Shaw, 2017; Batista & Leclercq, 2020; Li & Hensher, 2020; Pan & Zuo, 2020) dynamic programming (Bell, 1995), random regret minimization theory (Bekhor, Chorus, & Toledo, 2012; Chorus, 2012; Mai, Bastin, & Frejinger, 2017), the mental representation items (Arentze & Timmermans, 2012; Kazagli, Bierlaire, & Flötteröd, 2016; Kazagli, Bierlaire, & de Lapparent, 2020). However, the random utility models (RUMs; Manski, 1977; Ben-Akiva & Lerman, 1985), which assume users to choose according to a maximization utility rule, represents the most common tool for modelling route choice behaviour in transport planning: the paper restricts its attention to the models consistent with this hypothesis.

*Definition of route utility*. In principle, the definition of route utilities (actually represented in most cases by negative impedances) follows the usual specification of the perceived utility in RUMs, that is generally the sum of a systematic (i.e., deterministic) term and a random term. Other assumptions are represented by multiplicative (Kitthamkesorn & Chen, 2014, 2017; Kitthamkesorn, Chen, & Xu, 2015; Sharifi, Chen, Kitthamkesorn, & Song, 2015) or q-generalized (Nakayama & Chikaraishi, 2015a,b) relationships among deterministic and random terms of the utilities. In any case, the perceived route utility is simply considered as the negative of the perceived route impedance, in turn assumed as the sum of the impedances of its links.

*Substitution patterns amongst routes*. An unanimously acknowledged peculiarity of the route choice context is the need to account for the effects of topological overlapping of routes (i.e., choice alternatives) in a transport network, or at least in a sub-network as proposed by (Frejinger & Bierlaire, 2007). This aspect can be handled in two ways. The former is the approximated yet simple, and thus very popular, approach of specifying correction/penalty factors in the systematic utility of a MNL model, as a trick to mimic the effect of routes' overlapping on users' choice behaviour, i.e. that of reducing the probability of choosing each one of the overlapped routes. Relevant examples are the C-Logit model by (Cascetta, Nuzzolo, Russo, & Vitetta, 1996; Russo & Vitetta, 2003), and the Path size Logit model by (Ramming, 2002; Ben-Akiva & Bierlaire, 1999; Hoogendoorn-Lanser, 2005; Bekhor, Toledo, & Reznikova, 2009; Frejinger, Bierlaire, & Ben-Akiva, 2009; Hood, Sall, & Charlton, 2011). However, these models are known to fail in capturing the effects of route correlations on choice probabilities, as addressed with counterexamples (e.g. the well-known short-bypass long-bypass test network) by Prashker & Bekhor, (1998), Prashker & Bekhor (2004), Marzano (2006), Papola & Marzano (2013), Papola et al. (2018). The latter is a theoretically sounder approach that assumes appropriate distributional assumptions on the random part of route impedances to capture their correlations, see e.g. Cascetta (2009), Train (2009), . In fact, the Multinomial Probit (MNP) model has been for long time the sole model allowing for direct specification of a correlation matrix proportional to route overlapping (Daganzo & Sheffi, 1977), at the price of a non-closed-form probability statement. Alternative closed-form models with flexible correlations, namely the Link-Nested Logit (LNL) model and the Paired Combinatorial Logit (PCL) models, have been proposed, however without calculation of the underlying correlations. The latter has become practically feasible thanks to Abbe, Bierlaire, & Toledo (2007), Marzano & Papola (2008), Marzano, Papola, Simonelli, & Vitillo (2013) , Marzano (2014). Only recently, the Combination of Nested Logit (CoNL) model proposed by (Papola, 2016) and applied to route choice by (Papola et al., 2018) appeared as a viable alternative to the MNP model, thanks to the availability of a closed-form statement for both choice probability and correlations: this enables specification of its structural parameters to target any given covariance matrices.

This paper will investigate models consistent with the second approach: Section 2 will provide an exhaustive literature review.

*Enumeration of routes.* In terms of treatment of alternatives, route choice can be modelled following two approaches. The former is based on enumerating explicitly a route choice set by adopting well-established procedures (e.g. Prato, 2009) and, then, inferring on correlations amongst routes within the choice set based on the estimation of a RUM allowing for general correlation patterns. The latter seeks route choice probabilities without explicit enumeration of the routes, i.e., with a so-called implicit approach, a natural in most of standard traffic assignment procedures embedded also in commercial software.

Overall, regardless of the peculiar approaches described above, link flows provided by STA procedures reflect choice probabilities of their underlying route choice models: as a consequence, models handling properly correlation amongst routes are crucial to obtain realistic traffic flows. In addition, STA procedures should possibly avoid integral simulation for computing route choice probabilities and link flows.

This motivates the research presented in this paper, which aims to compute true correlations underlying closed-form route choice models consistent with the RUM paradigm. The remaining of the paper is organized as follows: Section 2 reports on the literature review concerning the route choice models assuming the above setup; Section 3 recalls route choice model formulations analysed in the paper, showing in particular the new proposed formulations; Section 4 show results on laboratory experiments; Section 5 draws conclusions and research prospects.

## 2. Literature review and paper notation

Daganzo & Sheffi (1977) pioneered the research on the correlation structure of utilities in a route choice model, by assuming correlation of a pair of route utilities to be proportional to their link-based topological overlapping, measured using a given link impedance. Such assumption can be applied to the network as a whole or, alternatively, only to a portion of the network given by primary, most likely perceived, roads, as proposed by Frejinger & Bierlaire (2007). Mathematically, let $K_{od}$ be a set of acyclic routes connecting the pair of centroids $o$ and $d$. Each route $k \in K_{od}$ is associated with an ordered set of links $L_k$. Moreover, letting $c_l$ be an additive link impedance and $C_k = \Sigma_l\ c_l\ \forall l \in L_k$, the corresponding route impedance, the assumption by Daganzo & Sheffi (1977) states that the covariance between random terms of a pair of routes $k$ and $k'$ is proportional to their topological overlapping, that is:

$$\sigma_{kk'} = Cov[\varepsilon_k, \varepsilon_{k'}] = \xi \cdot \sum_{l \in L_k \cap L_{k'}} c_l \qquad \forall k, k' \neq k \in K_{od} \quad (1)$$

where $\xi$ is the constant of proportionality and the variance of a random term of each route $k$ is proportional to its impedance:

$$\sigma_{kk} = Var[\varepsilon_k] = \xi \cdot \sum_{l \in L_k} c_l = \xi \cdot C_k \qquad \forall k \in K_{od} \quad (2)$$

that is a special case of (1) with $k=k'$. In turn, the generic correlation corresponding to (1) is given by:

$$\rho_{kk'} = \frac{\sigma_{kk'}}{\sqrt{\sigma_{kk} \sigma_{k'k'}}} = \frac{1}{\sqrt{C_k C_{k'}}} \cdot \sum_{l \in L_k \cap L_{k'}} c_l \qquad \forall k, k' \neq k \in K_{od} \quad (3)$$

The above assumption enables specifying directly a Multinomial Probit model (MNP) route choice model, as described in Section 3.1. Unfortunately, the MNP lacks a closed-form probability statement, leading to computational issues related to the need to simulate choice probabilities (Horowitz, Sparmann, & Daganzo, 1982; McFadden, 1989; Bunch, 1991; Geweke, 1991; Horowitz, 1991; Hajivassiliou, McFadden, & Ruud, 1996; Train, 2009, Bhat, 2011; Samimi, Mohammadian, Kawamura, & Pourabdollahi, 2014; Ding, Wang, Yang, Liu, & Lin, 2016; Patil et al., 2017). The same also apply to Mixed Logit (MMNL) applications to route choice, e.g. (Bekhor, Ben-Akiva, & Ramming, 2002; Frejinger & Bierlaire, 2007). Thus, a challenging research question arose in the literature, tackled by many researchers: can closed-form RUM-based route choice models be specified consistent with the Daganzo & Sheffi (1977) assumption?

The Generalized Extreme Value (GEV) modelling framework by McFadden (1978) represents the most straightforward mathematical vehicle in the attempt to tackle this question. In accordance with Section 1, GEV models allowing for fairly general (e.g. full) correlation matrix structures are a natural for route choice modelling, thus interest has been focussed primarily on the Cross-Nested Logit model (CNL; Vovsha, 1997) and on the Network GEV model (NGEV; Daly, 2001; Daly & Bierlaire, 2006; Newman, 2008). In this context, many models have been proposed so far, including the Link-Nested Logit (LNL) model by Vovsha & Bekhor (1998), the Paired-Combinatorial Logit (PCL) model by Prashker & Bekhor (1999), Pravinvongvuth & Chen (2005), Haghani, Shahhoseini, & Sarvi (2016). The CNL-based models are not the sole closed-form models to apply to route choice modelling. Castillo, Menéndez, Jiménez, & Rivas (2008) proposed a route choice Multinomial Weibit model (MNW), which assumes route impedances distributed as independent Weibull variables, interpretable also under the multiplicative random utility framework (see Fosgerau & Bierlaire, 2009). The MNW allows for perceived impedances to be heteroscedastic yet independent, thus not enabling correlation patterns like expression (3). Thus, Kitthamkesorn & Chen (2013, 2014), Kitthamkesorn et al. (2015), Xu, Chen, Kitthamkesorn, Yang, & Lo (2015) applied the MNW model to route choice with the addition of a path size attribute. However, as emphasized for the Path Size Logit, these models attempt to explain the effects of route overlapping on choice probabilities by means of attributes correcting the deterministic part of route utilities, and not by specifying proper correlation structures on the random terms.

Currently, there is no systematic assessment in the literature on the capability of GEV route choice models – namely the LNL and the PCL – to target also correlations (in addition to route choice probabilities) of the MNP model induced by (1)-(2). In fact, this is difficult task for GEV-based route choice models, because of their underlying non-closed form expression of the correlations, as extensively studied by Abbe et al. (2007), Marzano & Papola (2008), who showed that numerical evaluation of strongly nonlinear double-integrals is needed. Interestingly, Marzano et al. (2013) for the CNL model and Marzano (2014) for any GEV models, provided a simpler and more effective methodology for the calculation of GEV correlations, based on the numerical integration of a mono-dimensional integral, thus with parsimonious calculation times.

Another interesting closed-form route choice model is the Combination of Nested Logit (CoNL) model, recently proposed by Papola et al. (2018), which is a particular specification of the Combination of RUM (CoRUM) model proposed by Papola (2016), operationalized by means of an algorithm providing CoNL route choice probabilities on a set of explicitly enumerated routes for a given o-d pair. The CoNL combines Nested Logit (NL; Williams, 1977; Daly & Zachary, 1978; McFadden, 1978) models, characterized by block-diagonal correlation matrices as a consequence of partitioning the choice set into mutually exclusive nests, yielding, as demonstrated by Papola (2016), very general correlation patterns. A key feature of the CoNL is the availability of a closed-form statement for both choice probabilities and correlations. This distinct property of the CoNL allows handling effectively the relationship between the model specification (i.e. model structure, parameters) and its underlying correlations, thus enabling the specification of a CoNL route choice model targeting (Daganzo & Sheffi, 1977) correlations. Yet, the CoNL model is still homoscedastic (i.e. utilities/impedances of the alternatives have same variance) and, in addition, a full assessment of the correlation values arising by the specific setting of its structural parameters has not been exploited yet. Furthermore, it is worth exploring whether more effective specifications rather than that proposed by Papola et al. (2018) can be operationalized to improve CoNL route choice model performance.

Given these premises, this paper aims at providing a comprehensive analysis of concerned specifications and performance of LNL, PCL and CoNL route choice models. Specifically, the following research questions are investigated:

- analysing correlations underlying GEV route choice models with full correlation patterns;
- generalising the analysis of the properties of the CoNL route choice model proposed by Papola et al. (2018) on small-medium size networks and Tinessa, Marzano, Papola, Montanino, & Simonelli (2019) on a real world network, by means of a comparison that includes GEV and MNP route choice models;
- testing novel and more effective specifications of the structural parameters of GEV and CoNL models, including fixing a lower bound $\delta_{min}$ for the nesting parameters of LNL and CoNL models.

Since the final goal is evaluating the performance of the models both in terms of choice probabilities and in terms of correlations, the models have been tested on small-medium size networks, to better visualize and assess the performance of the models.

## 3. Route choice models under analysis: specification and correlation structure

### 3.1 The Probit model (MNP)

An MNP route choice model consistent with the assumptions (1)-(3) is simply obtained by imposing perceived link impedances to be randomly distributed as follows:

$$c_l^r \sim N(c_l, \xi \cdot c_l) \qquad (4)$$

The assumption (4), in the light of the properties of the Normal random variable, leads to a MNP route choice model with the overall error component structure by Daganzo & Sheffi (1977), that is assumptions (1)-(3). The parameter $\xi$ can be either estimated or exogenously determined based on prior assumption on the coefficient of variation of perceived link impedances. As recalled in the introduction, the MNP route choice model exhibit optimal performance in a variety of test networks applied in the literature, and thus is considered a benchmark for route choice modelling. Unfortunately, no closed-form probability statement exists for the MNP, thus numerical simulation[1] is required to get route choice probabilities.

### 3.2 CNL-based route choice models

#### 3.2.1 Model specifications

The CNL model is a GEV model (McFadden, 1978), obtained from the following generating function $G$ (see (Papola, 2004)):

$$G = \sum_{l \in L} \left[ \sum_{k \in l} \alpha_{kl}^{1/\delta_l} \cdot \exp\left(\frac{-C_k}{\theta_l}\right) \right]^{\delta_l} \qquad (5)$$

---

[1] In the remaining of the paper, unless otherwise specified, MNP choice probabilities are simulated based on $10^6$ Monte-Carlo draws, enough to get them stabilized in the tested networks.

being $\theta_l$ the variance parameter of the group $l$, $\delta_l=\theta_l/\theta_0$ the nesting parameter of the generic group $l$ ($\theta_0$ is the overall model variance parameter), and $\alpha_{lk}$ the inclusion coefficient (or membership degree) of alternative $k$ to nest $l$. Overall, (5) yields the following choice probability for the generic alternative $k$:

$$p(k) = \sum_{l' \in L} \frac{\alpha_{kl}^{1/\delta_l} \cdot \exp(-C_k/\theta_l)}{\sum_{k' \in l} \alpha_{k'l}^{1/\delta_l} \cdot \exp(-C_{k'}/\theta_l)} \cdot \frac{\left(\sum_{k' \in l} \alpha_{k'l}^{1/\delta_l} \cdot \exp(-C_{k'}/\theta_l)\right)^{\delta_l}}{\sum_{l' \in L}\left(\sum_{k' \in l'} \alpha_{k'l'}^{1/\delta_{l'}} \cdot \exp(-C_{k'}/\theta_{l'})\right)^{\delta_{l'}}} \quad (6)$$

A first application of (6) to route choice is the Link Nested Logit model (LNL) proposed by Vovsha & Bekhor (1998), that assumes a CNL structure with as many nests as the links of the network, each including all routes sharing the link representing that nest. Consistently, the inclusion coefficient is pre-specified (i.e. not estimated) as follows:

$$\alpha_{kl} = \frac{c_l}{C_k} \cdot a_{lk} \quad (7)$$

being $a_{lk}$ the generic entry of the link-route incidence matrix, i.e. $a_{lk}=1$ if $l \in k$ and 0 otherwise. It is easy to verify that (7) satisfies the CNL normalization constraint, see for example Abbe et al. (2007), Marzano & Papola (2008). Nesting parameters of the LNL can be specified in several ways:
- Vovsha & Bekhor (1998) assumed null nesting parameters, i.e. deterministic choice within each nest, which allows in turn the implementation of an LNL with implicit route enumeration;
- Bekhor & Prashker (2001) expressed nesting parameters as a function of the arithmetic mean of the inclusion parameters $\alpha_{kl}$ of all routes embedding $l$:

$$\delta_l = 1 - \frac{1}{n_l} \sum_{k \in K_{od}} \alpha_{kl} = 1 - \frac{1}{n_l} \sum_{k \in K_{od}} \frac{c_l}{C_k} \cdot a_{lk} \quad (8)$$

being $n_l$ the cardinality of the routes $k \in K_{od}$ that share the link $l$;
- Marzano (2006) expressed nesting parameters as a function of the geometric mean of the inclusion parameters $\alpha_{kl}$ of all routes embedding $l$:

$$\delta_l = 1 - \sqrt[n_l]{\prod_{k \in K_{od}} \alpha_{kl}} = 1 - \sqrt[2 \cdot n_l]{\prod_{k \in K_{od}} \frac{c_l}{C_k} \cdot a_{lk}} \quad (9)$$

This paper proposes and analyses the three formulations above, by introducing a positive lower-bound for all the nesting parameters. Indeed, albeit allowing implicit route enumeration, null nesting parameters lead to deterministic choices within nests, yielding wrong choice probabilities in test networks, see for example Prashker & Bekhor (2004), Papola et al. (2018).

Another interesting particularization of the CNL is the PCL specification (see Chu, 1989; Koppelman & Wen, 2000; Pravinvongvuth & Chen, 2005), which includes as many nests as the number of feasible pairs of alternatives, yielding the following generating function:

$$G = \sum_{k=1}^{n_k-1} \sum_{k'=k+1}^{n_k} (1-\sigma_{kk'}) \cdot \left\{ \exp\left[\frac{-C_k}{(1-\sigma_{kk'})}\right] + \exp\left[\frac{-C_{k'}}{(1-\sigma_{kk'})}\right] \right\}^{\frac{1-\sigma_{kk'}}{\theta_0}} \quad (10)$$

wherein $\sigma_{kk'}$ is the similarity parameter between alternatives $k$ and $k'$, and $\theta_0$ the model variance parameter. Resulting route probabilities are given by:

$$p(k) = \sum_{k' \neq k} \frac{\exp\left[\frac{-C_k}{(1-\sigma_{kk'})}\right]}{\exp\left[\frac{-C_k}{(1-\sigma_{kk'})}\right] + \exp\left[\frac{-C_{k'}}{(1-\sigma_{kk'})}\right]} \cdot \frac{\left\{\exp\left[\frac{-C_k}{(1-\sigma_{kk'})}\right] + \exp\left[\frac{-C_{k'}}{(1-\sigma_{kk'})}\right]\right\}^{\frac{1-\sigma_{kk'}}{\theta_0}}}{\sum_{k=1}^{n_k-1} \sum_{k'=k+1}^{n_k} \left\{\exp\left[\frac{-C_k}{(1-\sigma_{kk'})}\right] + \exp\left[\frac{-C_{k'}}{(1-\sigma_{kk'})}\right]\right\}^{\frac{1-\sigma_{kk'}}{\theta_0}}} \quad (11)$$

In a route choice context, $k$ and $k'$ represent two generic routes, and the similarity parameter $\sigma_{kk'}$ (see Gliebe, Koppelman, & Ziliaskopoulos, 1999) can be expressed as:

$$\sigma_{kk'} = \frac{C_{kk'}}{C_k + C_{k'} - C_{kk'}} \quad (12)$$

### 3.2.2 Covariances of LNL and PCL route choice models

The most effective method to calculate correlations underlying any GEV models, proposed by Marzano (2014), implies calculation of a one-dimensional integral, whose integrand function is available in closed form as a function of the generating function underlying the GEV model. This integral can be simulated very easily with parsimonious computational effort. Adapting the notation by Marzano (2014) to route choice, given a GEV model with $n_k$ random terms $\varepsilon_k$, generated by the function $G(y_{k_1},......,y_{k_{nk}})$ homogeneous of degree $\mu=1/\theta_0$ (where $\theta_0$ is the overall model variance parameter), the covariance between any pair of random terms $\varepsilon_k$ and $\varepsilon_{k'}$ is given by the following expression:

$$Cov[\varepsilon_k, \varepsilon_{k'}] = \frac{\pi^2 \theta_0^2}{6} - \frac{\theta_0}{2} \int_{-\infty}^{+\infty} x^2 \frac{d}{dx} \frac{G_k(0,...,e^{-x},...,1,...,0)}{G(0,...,e^{-x},...,1,...,0)} dx + \frac{1}{2}\left(E[\varepsilon_k] - E[\varepsilon_{k'}]\right)^2 \quad (13)$$

being $G_k(\cdot)$ the first partial derivative of the generating function $G(\cdot)$ with respect to its $k$-th component, and with both $G(\cdot)$ and $G_k(\cdot)$ calculated at a vector with all zero components but $e^{-x}$ as the $k$-th component and 1 as $k'$-th component.

## 3.3 The Combination of Nested Logit model (CoNL)

### 3.3.1 Mathematical formulation

The CoNL model is a particular instance of the CoRUM proposed by Papola (2016). The CoNL is an additive RUM, whose underlying cumulative distribution function (cdf) is the linear combination of a set I of $n_{MC}$ different cdf's of NL (Williams, 1977; Daly & Zachary, 1978; McFadden, 1978) models with the same $n$ random terms $(\varepsilon_{k_1},...,\varepsilon_{k_{nk}})$, with $n_{MC}$ non-negative weights $w^1,..., w^{nMC}$ summing up to 1, that is:

$$F_{CoNL}(\varepsilon_{k_1},...,\varepsilon_{k_{nk}}) = \sum_{i \in I} w^i \cdot F_{NL}^i(\varepsilon_{k_1},...,\varepsilon_{k_{nk}}) = \sum_{i \in I} w^i \cdot \exp- \sum_{l^i} \left(\sum_{k' \in l^i} e^{-\varepsilon_{k'}/\theta_l^i}\right)^{\delta_l^i} \quad (14)$$

$$\sum_{i \in I} w^i = 1 \quad (15)$$

being $l^i$ the generic nest associated with the $i$-th mixing NL, $k'$ the generic alternative belonging to such nest, and $\delta_l^i = \theta_l^i / \theta_o$ the nesting parameter associated to nest $l^i$. In turn, the probability statement and the variances/covariances of the CoNL are convex combinations of the corresponding expressions of the mixing NL components:

$$p(k) = \sum_{i \in I} w^i \cdot p_{NL}^i(k) = \sum_{i \in I} w^i \cdot \frac{e^{-C_k/\theta_{l(k)}^i} \cdot \left(\sum_{k' \in l^i(k)} e^{-C_{k'}/\theta_{l(k)}^i}\right)^{\delta_{l(k)}^i - 1}}{\sum_{l^i} \left(\sum_{k' \in l^i} e^{-C_{k'}/\theta_{l^i}^i}\right)^{\delta_{l^i}^i}} \tag{16}$$

$$Var[\varepsilon_k] = \sum_{i \in I} w^i \cdot Var^i[\varepsilon_k] = \sum_{i \in I} w^i \cdot \pi^2/6 = \pi^2/6 \tag{17}$$

$$Cov[\varepsilon_k \varepsilon_{k'}] = \sum_{i \in I} w^i \cdot Cov^i[\varepsilon_k \varepsilon_{k'}] = \pi^2/6 \cdot \sum_{i \in I^{kk'}} w^i \cdot [1 - (\delta_{l(k,k')}^i)^2] \quad \forall k, k' \tag{18}$$

being $l^i(k)$ the specific nest $l$ of the $i$-th component containing alternative $k$ and $I^{kk'}$ the subset of all NL mixing components exhibiting a nest $l(k,k')$ including both alternatives $k$ and $k'$. The corresponding general expression of the CoNL correlations is:

$$\rho_{kk'} = \frac{Cov[\varepsilon_k \varepsilon_{k'}]}{(Var[\varepsilon_k])^{0.5} \cdot (Var[\varepsilon_{k'}])^{0.5}} = \sum_{i \in I^{kk'}} w^i \cdot [1 - (\delta_{l(k,k')}^i)^2] \quad \forall k, k' \tag{19}$$

The operationalization of the CoNL depends on the definition of the NL mixing components and the computation of its structural parameters ($w$'s and $\delta$'s), described in the following sub-section.

### 3.3.2 Application of the CoNL to route choice

The CoNL model has been particularized to route choice by Papola et al. (2018) and applied to a real world network by Tinessa et al. (2019), leveraging the concept of "shared links" – that is, links embedded in multiple routes –such that a shared link defines a nest including all routes embedding that link. Papola et al. (2018) defined the structure of the NL mixing components and a consistent specification of the structural parameters of the CoNL, able to target the correlations (3); in particular, the proposed CoNL route choice specification is such that each nest representing a shared link can be included in multiple mixing components. Notably, the concept of shared link underlies also the specification of the LNL model, however in a way that each nest representing a shared link is present only once in the LNL model, yielding a CNL structure implied by routes belonging to all nests representing their shared links.

The CoNL route choice model was operationalised by Papola et al. (2018) by means of a recursive algorithm that searches for a CoNL structure compliant with the following requirements:

*(a)* each efficient route[2] (see Dial, 1971) belongs to a unique nest within each mixing component, consistent with the definition of NL;
*(b)* each mixing component embeds all the efficient routes connecting that o-d pair, consistent with the CoNL assumption (14).

The above conditions imply that a mixing component may embed also nests/links including only one route and allows for the same nest/link may to appear in more than one mixing component, consistent with the above. Notably, if the number $n_{sl}^{i,od}$ of nests/links of a mixing component $i$ equals the total number $n_k^{od}$ of (efficient) routes connecting a generic o-d pair, there is only one alternative within each nest, thus that NL mixing component collapses into a simple MNL.

Overall, the CoNL route choice model structure by Papola et al. (2018) implies (19) to become:

$$\rho_{kk'} = \sum_{i \in I^{kk'}} w^i \cdot [1 - (\delta_{l(k,k')}^i)^2] = \sum_{l \in L_k \cap L_{k'}} \sum_{i \in I^l} w^i \cdot (1 - \delta_l^{i2}) \tag{20}$$

---

[2] An efficient route with reference to a the origin $o$ is a route composed of efficient links with reference to the same origin, i.e. links $(i,j)$ verifying the inequality $C_{oi} < C_{oj}$, being the latter the minimum route cost to reach, respectively, the node $i, j$ from $o$.

where, $I^l$ denotes the subset of mixing components which a link $l$ belongs to and $I^{kk'} \equiv \{\cup_l I^l, l \in L_k \cap L_{k'}\}$. Letting $\delta_l^i = \delta_l$, $\forall i \in I^l, \forall l \in L_k \cap L_{k'}$, $\forall k, k' \neq k \in K_{od}$ within (19), it occurs:

$$\rho_{kk'} = \sum_{l \in L_k \cap L_{k'}} (1 - \delta_l^2) \cdot \sum_{i \in I^l} w^i \tag{21}$$

The covariances by Daganzo & Sheffi (1977) can be targeted by contrasting (21) and (3), obtaining the following system of equations:

$$(1 - \delta_l^2) \cdot \sum_{i \in I^l} w^i = \frac{c_l}{C_{od,\min}} \quad \forall l \in L_k \cap L_{k'}, \quad \forall k, k' \neq k \in K_{od} \tag{22}$$

obtained by assuming $C_k = C_{k'} = C_{od,\min}$ $\forall k, k' \neq k \in K_{od}$ in (3). Notice that, according to (2), the latter assumption is consistent with the hypothesis of homogenous variance (homoscedasticity) of the CoNL.

The system of equations (22) consists of $n_{sl}^{od}$ equations (as many as the shared links within a given o-d pair) in $n_{sl}^{od}$ (as many as the $\delta$s) + $n_{MC}$ (as many as the $w$'s) unknowns. Its solution is not straightforward and even unfeasible, because of the need to satisfy the constraint (15) and the [0-1] bound of the $\delta$s. The reader may refer to Papola et al. (2018), Tinessa et al. (2019) for an in-depth investigation of this issue. Overall, a fairly general specification of the generic $w_i$ $\forall i \in 1 \ldots n_{MC}$, embedding those proposed by Papola et al. (2018), Tinessa et al. (2019), is the following:

$$w^i = \frac{e_i \cdot \left[f(\mathbf{c}^i)\right]^\gamma}{\sum_{i' \in I} e_{i'} \cdot \left[f(\mathbf{c}^{i'})\right]^\gamma} \text{ with } e_{i'} = \begin{cases} 1 & \text{if } 1 < n_l^{i',od} < n_k^{od} \\ 0 & \text{otherwise} \end{cases} \tag{23}$$

wherein $f(\mathbf{c}^{i'})$ is a function of the impedances of the $n_{sl}^{i',od}$ shared links[3] belonging to the generic $i$–th mixing component, $\gamma$ a parameter to estimate and $e_{i'}$ a binary 1/0 value, indicating if $i'$ is a NL mixing component ($e_{i'} = 1$) or a simple MNL mixing component ($e_{i'} = 0$), to account for the aforementioned limiting case. In the remaining of the paper it will be assumed $\gamma = 1$.

Overall, four different specifications have been considered for $f(\cdot)$ in equation (23), as summarised in the following TABLE 1: notably, they are all original but (24), proposed by Papola et al. (2018).

**TABLE 1**: *CoNL route choice model – Possible formulations of $f(\mathbf{c}^i)$.*

| (24) | (25) | (26) | (27) |
|---|---|---|---|
| $f(\mathbf{c}^i) = E\{c_l^i\}$ | $f(\mathbf{c}^i) = E\left\{\dfrac{c_l^i}{n_l^{od}}\right\}$ | $f(\mathbf{c}^i) = \min_{l=1,\ldots,n_{sl}^{i,od}}\left\{\dfrac{c_l^i}{n_l^{od}}\right\}$ | $f(\mathbf{c}^i) = \max_{l=1,\ldots,n_{sl}^{i,od}}\left\{\dfrac{c_l^i}{n_l^{od}}\right\}$ |

wherein $n_l^{od}$ represents the number of times the link $l$ appears across mixing components for the o-d pair *od*.

The new proposed formulations (25)-(27) aim to achieve a more effective distribution of weights across mixing components, in the attempt to target correlations (3) more closely, splitting the overall contribution needed to target the correlation generated by a link across all $n_l^{od}$ mixing components including the nest representing that link. Relevant performance checks will be presented in Sections 4.3 and 4.4. Once defined the weights, nesting parameters are easily obtained from (22):

---

[3] Notably, in general it occurs $n_{sl}^i \leq n_l^i$, i.e. the number of "real" nests including at least two elemental alternatives is less or equal than the number of total nests included in a mixing component.

$$\delta_l = \begin{cases} \max\left\{\delta_{\min}; \sqrt{1 - \dfrac{c_l}{C_{od,\min} \cdot \sum_{i \in I^l} w^i}}\right\} & \text{if } 1 - \dfrac{c_l}{C_{od,\min} \cdot \sum_{i \in I^l} w^i} > 0 \\ \delta_{\min} & \text{otherwise} \end{cases} \quad (28)$$

wherein $\delta_{\min}$ is a lower bound for the nesting parameters. The presence of a lower bound for the nesting parameter value is behaviourally justifiable with the same arguments illustrated in Section 3.2 for the LNL.

## 4. Experimental analysis

This section contrasts the performance of the route choice models illustrated in Section 2 – plus the Multinomial Logit (MNL) model as reference point – in terms of choice probabilities and correlations, in several small- and medium-size networks. Consistent with the introduction, the target is represented by the MNP model: thus, the distance between target probabilities and correlations by MNP and by the tested model is reported in the following tables in terms of the mean of squared errors (MSE indicator) for both probabilities (MSE·$10^4$) and correlations (MSE·$10^3$). Notably, correlations are compared both in the space of the utilities (i.e. looking at the Full Correlation Matrix, FCM) and in the space of the utility differences (i.e. the Reduced Correlation Matrix, RCM). Following Bunch (1991), Train (2009), the latter can be obtained easily from the former by a linear transformation of the full correlation matrix. Finally, results are presented for different values of $\delta_{\min}$ for all concerned models, that is the CoNL and LNL models. Thus, the performance of the LNL models proposed in the literature corresponds to the case of $\delta_{\min}=0$.

The underlying explicit choice sets have been computed by drawing link impedance errors from monovariate Normal distribution with link impedance as average value (see Sheffi & Powell, 1982). All models are tested by assuming two different values, respectively 0.1 and 0.2, of the overall coefficient of variation (*cv*), which obviously impact on choice probabilities and covariances whilst correlations are kept constant. This means setting the overall variance parameter $\theta_0$ of both GEV and CoNL model such that (Cascetta et al., 1996; Papola & Marzano, 2013; Papola et al., 2018):

$$\theta_0 = \frac{\sqrt{6} \cdot cv \cdot C_{od,\min}}{\pi} \quad (29)$$

which corresponds to the following setting for the MNP parameter:

$$\xi = cv^2 \cdot C_{od,\min} \quad (30)$$

Four networks have been considered for the experimental analysis:

- the Braess' network (FIGURE 2), that is the simplest network characterized by a non-block-diagonal correlation matrix, tested on two different scenarios;
- two mesh networks (FIGURE 5 and FIGURE 8): the former is a 2x2 regular mesh with various specifications of link impedances, the latter is the same proposed by Papola et al. (2018);
- the Sioux-Falls network (FIGURE 10).

Notably, another network is usually presented in the literature to assess properties of route choice models, which is the four links-three paths network with a long bypass and a short bypass (FIGURE 1). Concerned results of route choice models presented in Section 2 on this test network are not reported here for the sake of brevity. However, it is important to remark that all modified Logit formulations (both C-Logit and Path Size Logit) fail to target expected probabilities (as recalled in Papola et al., 2018) in such network. In addition, when applied to this network, both LNL, PCL and CoNL collapse to a NL model, which is known to be capable to match expected choice probabilities and underlying correlations for any *h* on this network. Note that results are independent of the values assumed for *c*, *h* and *k*, because the additive utility maximization setup implies that only differences in utilities/impedances matter.

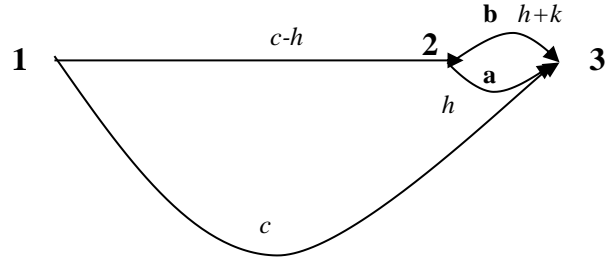

**FIGURE 1:** Four links / three routes (Daganzo and Sheffi, 1977) test network.

The following sub-sections report on results on the four above listed test networks.

## 4.1 Braess' network

Contrasted models are tested first on the Braess' network (FIGURE 2), with a single o-d pair **1-4** characterized by three acyclic routes and two equal values of Daganzo & Sheffi (1977) correlations. Notably, albeit rather simplistic, Braess' network allows testing route choice model performances and drawing informative conclusions; results are presented for *a*=4 and *b*=5 (that is *a/b*=0.8), in the cases *h*=0 and *h*=0.1.

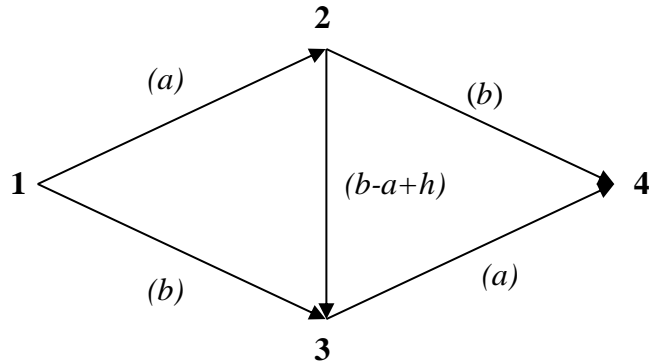

**FIGURE 2:** Braess' network.

Interestingly, the CoNL model yields closest results to MNP, both for correlations and probabilities, for $\delta_{min}<0.4$. The LNL with null $\delta$'s by Vovsha & Bekhor (1998) works only with *h*=0 (see TABLE 3), i.e. when equal impedance routes nullify the side effect of within-nest deterministic choice, and provides very unsatisfactory results – even worse than the MNL – as soon as route costs are slightly differentiated (e.g. with *h*=0.1), as shown in TABLE 4. In addition, a lower bound of 0.3÷0.4 on $\delta_{min}$ allows optimizing the performance of the LNL model, whilst larger values of $\delta_{min}$ tend to magnify the distance between LNL/CoNL and MNP correlations (TABLE 2, TABLE 3, TABLE 4, FIGURE 3 and FIGURE 4).

This can be explained because the higher is $\delta_{min}$ the narrower is the underlying domain of correlation matrices allowed by such models; overall, the benefit of setting an intermediate $\delta_{min}$ for the LNL confirms the findings by Papola et al. (2018) on the CoNL model. Finally, the PCL fails in targeting MNP correlations and probabilities. Additionally, FIGURE 3 and FIGURE 4 show also a consistency between the objectives of targeting correlations and probabilities. Indeed, as $\delta_{min}$ increases, not only the distances between LNL/CoNL and MNP correlations increase – for the reason mentioned above – but also those between LNL/CoNL and MNP probabilities. Overall, adopting full-correlation models in route choice modelling is a necessary condition.

TABLE 2: **Correlations in the Braess' network, for the o-d pair *1-4* (*k*=0).**

| | | $\rho_{AB}$ | | | | |
|---|---|---|---|---|---|---|
| $\delta_{min}$ | MNP | CoNL (any) | LNL ($\delta_l=\delta_{min}$) | LNL with (8) | LNL with (9) | PCL with (12) |
| 0 | | 0.44 | 0.42 | | | |
| 0.1 | | 0.44 | 0.42 | | | |
| 0.2 | 0.44 | 0.44 | 0.41 | 0.31 | 0.38 | 0.24 |
| 0.3 | | 0.44 | 0.39 | | | |
| 0.4 | | 0.42 | 0.36 | | | |

**TABLE 3:** Braess network, o-d pair *1-4* (*k*=0) – MSE values for the tested route choice models.

| | MNL | | | | CoNL (any) | | | | PCL with (12) | | | |
|---|---|---|---|---|---|---|---|---|---|---|---|---|
| | Correlations | | Choice prob. | | Correlations | | Choice prob. | | Correlations | | Choice prob. | |
| | FCM | RCM | $cv$=0.1 | $cv$=0.2 | FCM | RCM | $cv$=0.1 | $cv$=0.2 | FCM | RCM | $cv$=0.1 | $cv$=0.2 |
| $\delta_{min}$ | MSE (*10³) | | MSE (*10⁴) | | MSE (*10³) | | MSE (*10⁴) | | MSE (*10³) | | MSE (*10⁴) | |
| 0 | | | | | | | | | | | | |
| 0.1 | | | | | | | | | | | | |
| 0.2 | | | | | **0.00** | **0.00** | **0.85** | **1.30** | | | | |
| 0.3 | | | | | | | | | | | | |
| 0.4 | | | | | 0.27 | 0.10 | 1.75 | 2.37 | | | | |
| 0.5 | **87.79** | **14.59** | **22.86** | **24.97** | 2.14 | 0.74 | 3.69 | 4.57 | **19.15** | **4.85** | **6.10** | **14.93** |
| 0.6 | | | | | 6.88 | 2.08 | 6.31 | 7.45 | | | | |
| 0.7 | | | | | 15.95 | 4.19 | 9.58 | 10.97 | | | | |
| 0.8 | | | | | 31.08 | 7.04 | 13.45 | 15.09 | | | | |
| 0.9 | | | | | 54.27 | 10.55 | 17.89 | 19.77 | | | | |
| 1 | | | | | 87.79 | 14.59 | 22.86 | 24.97 | | | | |

| | LNL ($\delta_l=\delta_{min}$) | | | | LNL with (8) | | | | LNL with (9) | | | |
|---|---|---|---|---|---|---|---|---|---|---|---|---|
| | Correlations | | Choice prob. | | Correlations | | Choice prob. | | Correlations | | Choice prob. | |
| | FCM | RCM | $cv$=0.1 | $cv$=0.2 | FCM | RCM | $cv$=0.1 | $cv$=0.2 | FCM | RCM | $cv$=0.1 | $cv$=0.2 |
| $\delta_{min}$ | MSE (*10³) | | MSE (*10⁴) | | MSE (*10³) | | MSE (*10⁴) | | MSE (*10³) | | MSE (*10⁴) | |
| 0 | **0.85** | **0.32** | **0.00** | **0.05** | | | | | | | | |
| 0.1 | 0.98 | 0.35 | 0.10 | 0.28 | | | | | **2.89** | **0.97** | **2.16** | **2.85** |
| 0.2 | 1.42 | 0.50 | 0.65 | 1.05 | 10.48 | 2.99 | 6.68 | 7.84 | | | | |
| 0.3 | 2.39 | 0.81 | 1.70 | 2.31 | | | | | | | | |
| 0.4 | 4.29 | 1.38 | 3.26 | 4.09 | | | | | 4.29 | 1.38 | 3.26 | 4.09 |
| 0.5 | 7.70 | 2.29 | 5.32 | 6.36 | | | | | 7.70 | 2.29 | 5.32 | 6.36 |
| 0.6 | 13.49 | 3.66 | 7.88 | 9.14 | 13.49 | 3.66 | 7.88 | 9.14 | 13.49 | 3.66 | 7.88 | 9.14 |
| 0.7 | 22.77 | 5.55 | 10.93 | 12.41 | 22.77 | 5.55 | 10.93 | 12.41 | 22.77 | 5.55 | 10.93 | 12.41 |
| 0.8 | 36.99 | 8.01 | 14.45 | 16.15 | 36.99 | 8.01 | 14.45 | 16.15 | 36.99 | 8.01 | 14.45 | 16.15 |
| 0.9 | 57.94 | 11.04 | 18.44 | 20.34 | 57.94 | 11.04 | 18.44 | 20.34 | 57.94 | 11.04 | 18.44 | 20.34 |
| | 87.79 | 14.59 | 22.86 | 24.97 | 87.79 | 14.59 | 22.86 | 24.97 | 87.79 | 14.59 | 22.86 | 24.97 |

**TABLE 4:** Braess network, o-d pair *1-4* (k=0.1) – MSE values for the tested route choice models.

| | MNL | | | | CoNL (any) | | | | PCL with (12) | | | |
|---|---|---|---|---|---|---|---|---|---|---|---|---|
| | Correlations | | Choice prob. | | Correlations | | Choice prob. | | Correlations | | Choice prob. | |
| | FCM | RCM | $cv=0.1$ | $cv=0.2$ | FCM | RCM | $cv=0.1$ | $cv=0.2$ | FCM | RCM | $cv=0.1$ | $cv=0.2$ |
| $\delta_{min}$ | MSE ($*10^3$) | | MSE ($*10^4$) | | MSE ($*10^3$) | | MSE ($*10^4$) | | MSE ($*10^3$) | | MSE ($*10^4$) | |
| 0 | | | | | | | | | | | | |
| 0.1 | | | | | | | | | | | | |
| 0.2 | | | | | **0.00** | **0.01** | 0.92 | **0.01** | | | | |
| 0.3 | | | | | | | | | | | | |
| 0.4 | | | | | 0.22 | 0.06 | **0.00** | 0.63 | | | | |
| 0.5 | **86.83** | **14.03** | **28.17** | **27.02** | 1.99 | 0.61 | 1.46 | 2.88 | **18.94** | **4.57** | **10.04** | **11.23** |
| 0.6 | | | | | 6.61 | 1.87 | 4.84 | 6.23 | | | | |
| 0.7 | | | | | 15.54 | 3.89 | 9.45 | 10.43 | | | | |
| 0.8 | | | | | 30.51 | 6.65 | 14.98 | 15.35 | | | | |
| 0.9 | | | | | 53.51 | 10.07 | 21.25 | 20.90 | | | | |
| 1 | | | | | 86.83 | 14.03 | 28.17 | 27.02 | | | | |

| | LNL ($\delta_l=\delta_{min}$) | | | | LNL with (8) | | | | LNL with (9) | | | |
|---|---|---|---|---|---|---|---|---|---|---|---|---|
| | Correlations | | Choice prob. | | Correlations | | Choice prob. | | Correlations | | Choice prob. | |
| | FCM | RCM | $cv=0.1$ | $cv=0.2$ | FCM | RCM | $cv=0.1$ | $cv=0.2$ | FCM | RCM | $cv=0.1$ | $cv=0.2$ |
| $\delta_{min}$ | MSE ($*10^3$) | | MSE ($*10^4$) | | MSE ($*10^3$) | | MSE ($*10^4$) | | MSE ($*10^3$) | | MSE ($*10^4$) | |
| 0 | **0.85** | **0.27** | 142.32 | 102.53 | | | | | | | | |
| 0.1 | 0.97 | 0.30 | 52.36 | 18.37 | | | | | **2.86** | **0.88** | **0.02** | **0.54** |
| 0.2 | 1.41 | 0.44 | 7.77 | 1.21 | 10.52 | 2.83 | 5.02 | 6.47 | | | | |
| 0.3 | 2.37 | 0.73 | **0.43** | **0.13** | | | | | | | | |
| 0.4 | 4.25 | 1.25 | 0.49 | 1.79 | | | | | 4.25 | 1.25 | 0.49 | 1.79 |
| 0.5 | 7.63 | 2.12 | 2.99 | 4.54 | | | | | 7.63 | 2.12 | 2.99 | 4.54 |
| 0.6 | 13.35 | 3.43 | 6.69 | 7.99 | 13.35 | 3.43 | 6.69 | 7.99 | 13.35 | 3.43 | 6.69 | 7.99 |
| 0.7 | 22.53 | 5.25 | 11.19 | 12.01 | 22.53 | 5.25 | 11.19 | 12.01 | 22.53 | 5.25 | 11.19 | 12.01 |
| 0.8 | 36.59 | 7.63 | 16.32 | 16.55 | 36.59 | 7.63 | 16.32 | 16.55 | 36.59 | 7.63 | 16.32 | 16.55 |
| 0.9 | 57.31 | 10.57 | 22.00 | 21.56 | 57.31 | 10.57 | 22.00 | 21.56 | 57.31 | 10.57 | 22.00 | 21.56 |
| 1 | 86.83 | 14.03 | 28.17 | 27.02 | 86.83 | 14.03 | 28.17 | 27.02 | 86.83 | 14.03 | 28.17 | 27.02 |

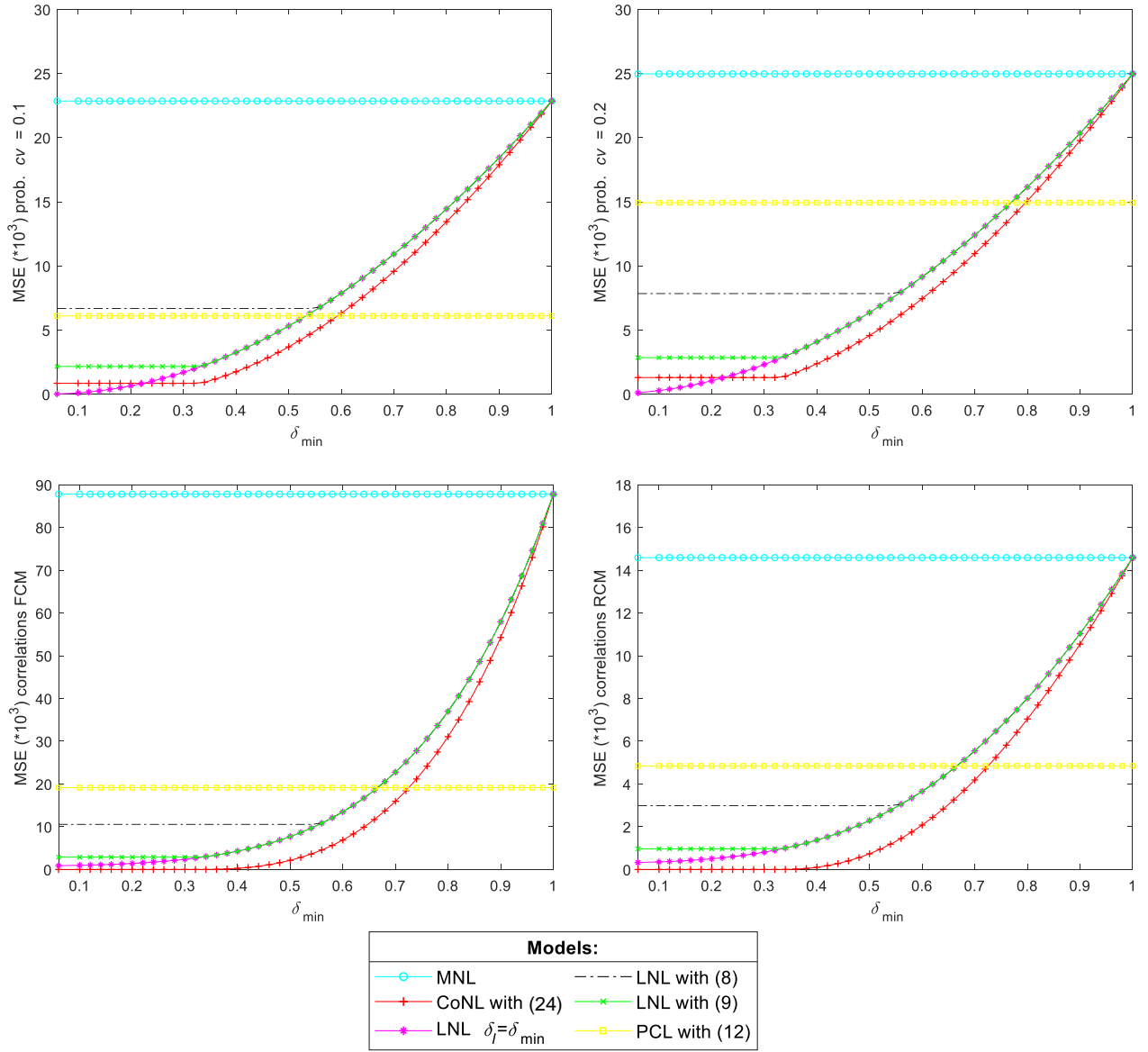

**FIGURE 3:** Braess network, o-d pair *1-4* (k=0) – Distance between tested models and MNP in terms of probabilities (cv=0.1 top left, cv=0.2 top right) and correlations (Full Correlation Matrix, bottom left, Reduced Correlation Matrix, bottom right).

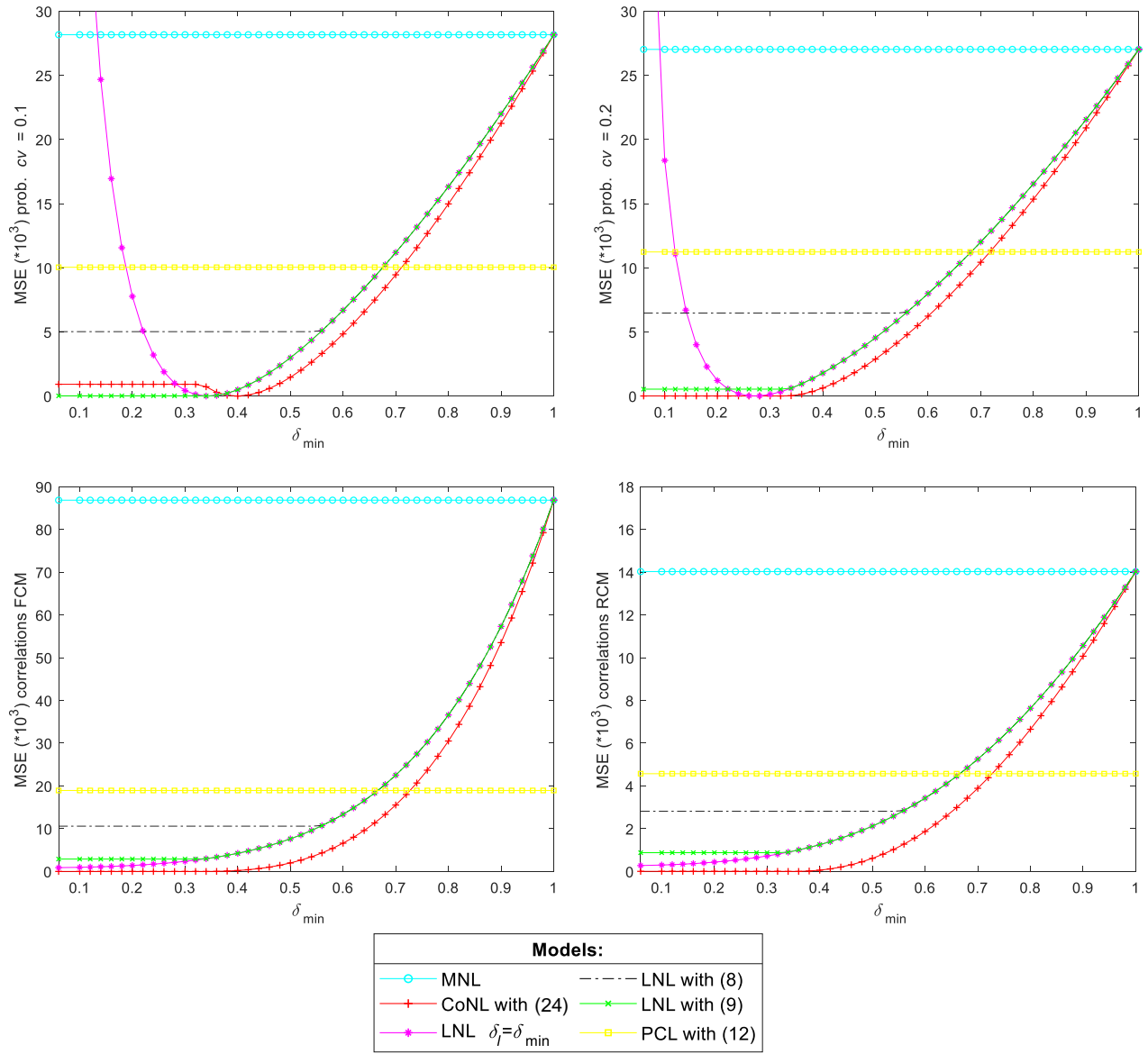

**FIGURE 4:** Braess network, o-d pair *1-4* (k=0.1) – Distance between tested models and MNP in terms of probabilities (cv=0.1 top left, cv=0.2 top right) and correlations (Full Correlation Matrix, bottom left, Reduced Correlation Matrix, bottom right).

## 4.2 Mesh network

The first tested mesh network is the two-way links mesh depicted in FIGURE 5, with a single o-d pair **1-9** and two different configurations of link impedances.

*Case #1 - Equal link impedances.* This impedance configuration leads, consistent with (3), to a homoscedastic covariance matrix and a choice set consisting of 6 efficient routes. Also, the particular setting and topology of the network is such that all specifications for the weights of the CoNL illustrated in Section 3.3 collapse to the same value, that is the inverse of the number of mixing components plus one.

Results obtained in this case are in line with the trends observed for the Braess' network. In particular, the CoNL targets best both MNP correlations and probabilities (see FIGURE 6 and TABLE 5 in the Appendix), and the role of $\delta_{min}$ is even more evident, yielding an optimal setting in between 0.2÷0.3, after which the distances between CoNL and MNP correlations and between CoNL and MNP probabilities start again to increase consistently (see Figure 6). In addition, the LNL needs a higher threshold for $\delta_{min}$, equal to 0.4, to exhibit optimal fitting of MNP correlations and probabilities, and the PCL fails again in targeting expected results.

*Case #2 - Different link impedances drawn from a uniform distribution.* This impedance configuration creates a correlation context inherently heteroscedastic, not consistent with the underlying assumptions of all GEV-based and the CoNL model. Notwithstanding, results are in line with what observed so far: the CoNL appears again the best model, even if for a particular combination of $\delta_{min}$ (0.5) and *cv* (0.2) the LNL yields best fitting of MNP probabilities (see FIGURE 7 and TABLE 6 in the appendix). Moreover, a greater $\delta_{min}$ with respect to the previous test networks is needed to optimize both CoNL and LNL performances in terms of probabilities.

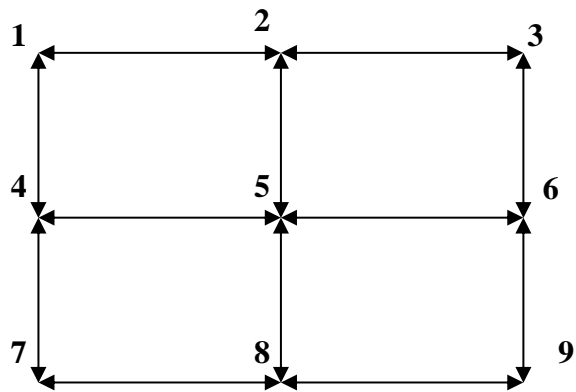

**FIGURE 5:** Regular 2 x 2 mesh network.

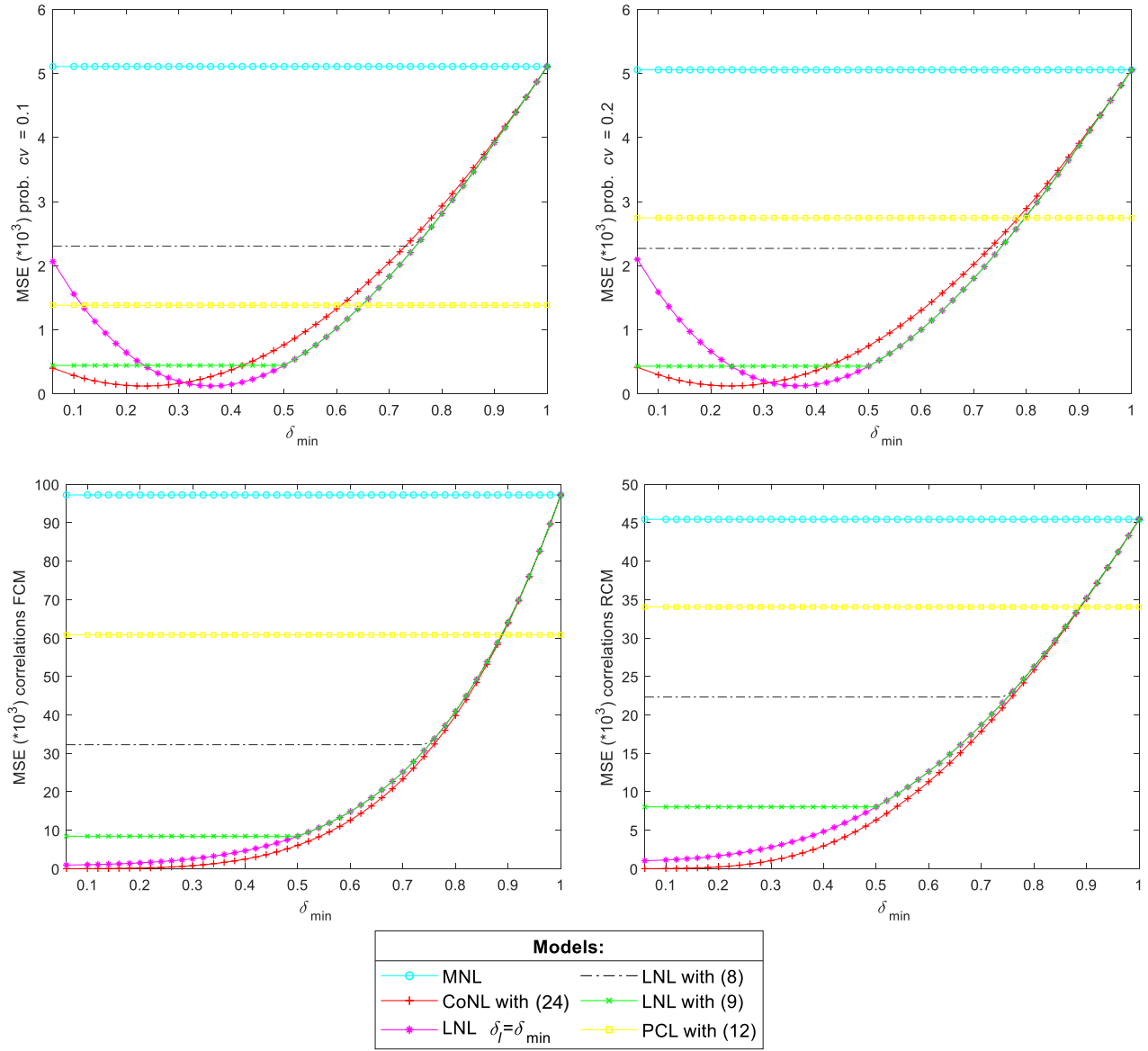

**FIGURE 6:** Regular mesh network, o-d pair *1-9,* case #1– Distance between tested models and MNP in terms of probabilities (cv=0.1 top left, cv=0.2 top right) and correlations (Full Correlation Matrix, bottom left, Reduced Correlation Matrix, bottom right).

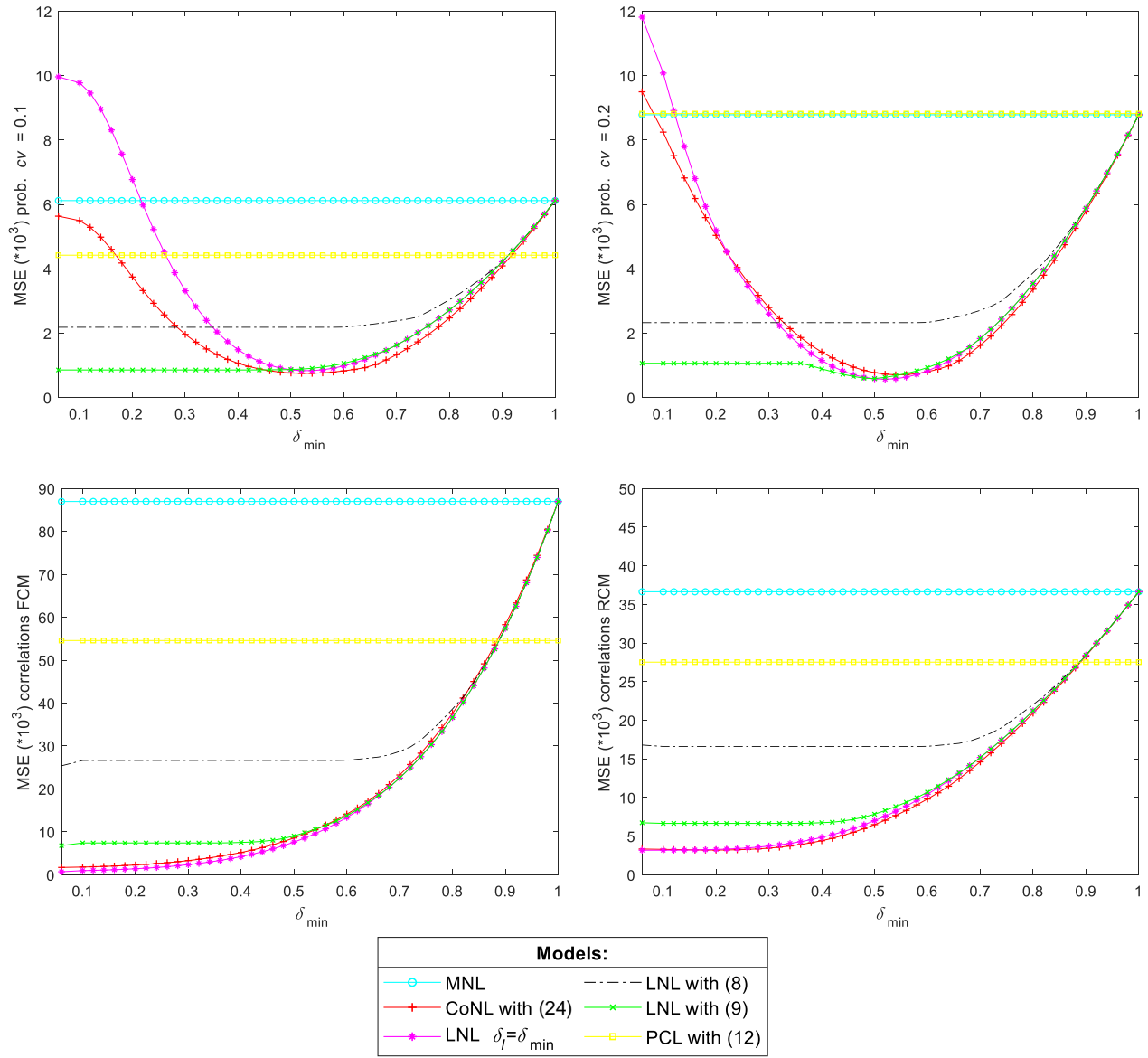

**FIGURE 7:** Regular mesh network, o-d pair **1-9**, case #2 – Distance between tested models and MNP in terms of probabilities (cv=0.1 top left, cv=0.2 top right) and correlations (Full Correlation Matrix, bottom left, Reduced Correlation Matrix, bottom right).

## 4.3 Mesh network with long bypass

The network depicted in FIGURE 8 represents an even more challenging correlation scenario for the CoNL model, as extensively illustrated by Papola et al. (2018), because of the presence of a long bypass (link *1-9*) that yields a particular configuration of its mixing components. The considered o-d pair **1-12** is characterized by 18 efficient routes, with similar route impedances and a slight heteroscedastic covariance matrix.

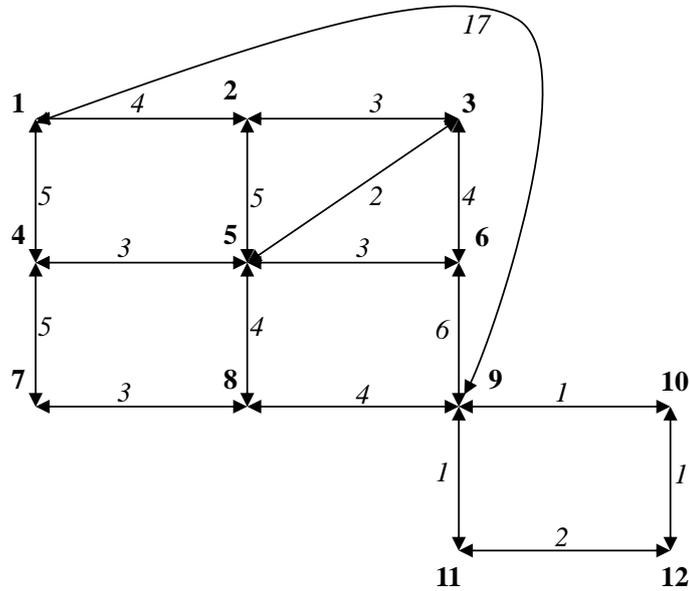

**FIGURE 8:** Mesh network: topology and link impedances.

CoNL performances are much more stable when varying $\delta_{min}$, even if the best fitting of MNP probabilities is reached by the LNL with $\delta_{min}$=0.4, slightly outperforming the CoNL that attains its best fitting for $\delta_{min}$=0.3 (see FIGURE 9 and TABLE 7 in the Appendix). Interestingly, the new proposed specifications of the weights of the CoNL (see Section 3.3) outperform the original specification by Papola et al. (2018) in terms of fitting both MNP correlations and MNP probabilities.

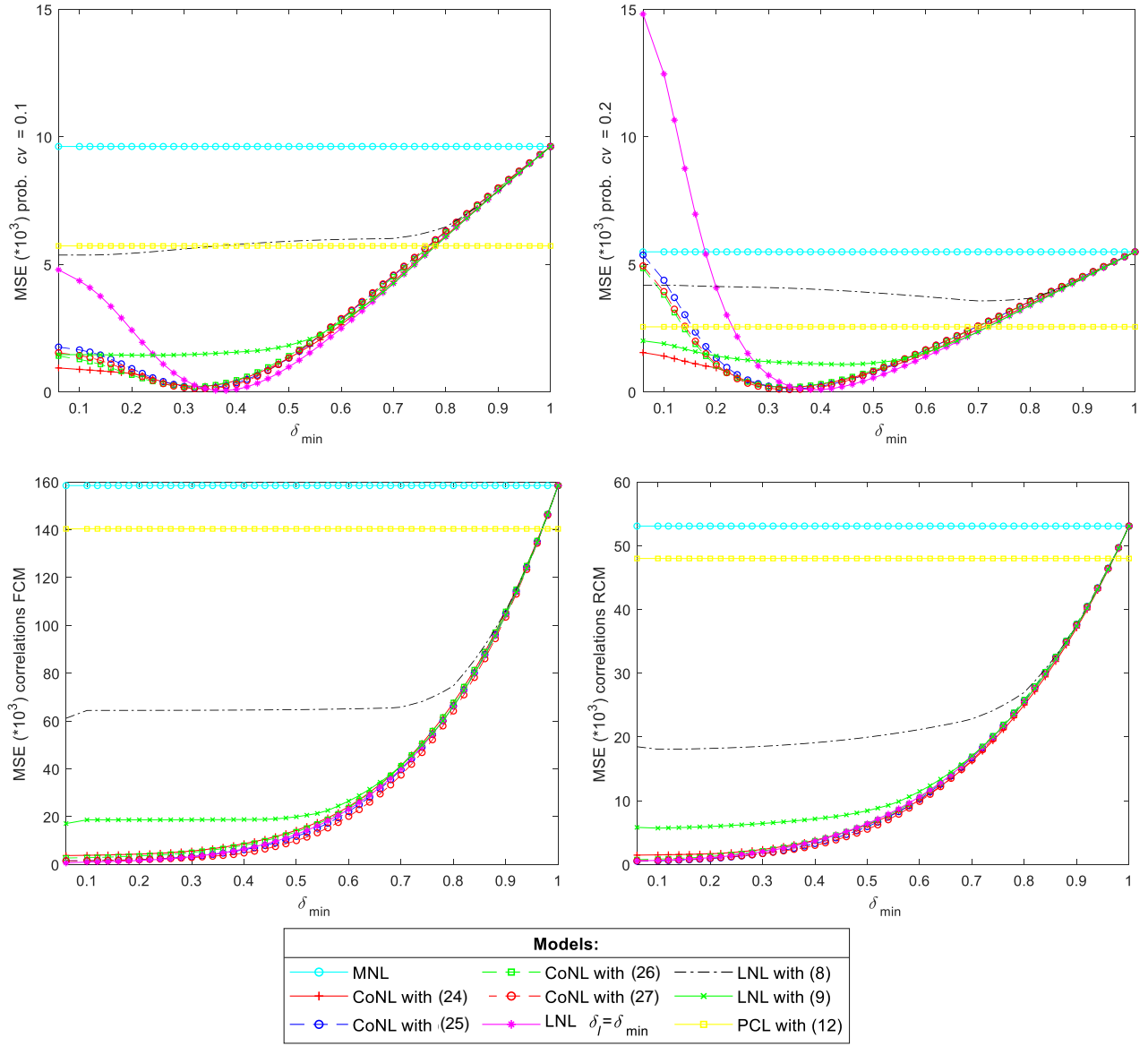

**FIGURE 9:** Mesh network with bypass, o-d pair *1-12* – Distance between tested models and MNP in terms of probabilities (cv=0.1 top left, cv=0.2 top right) and correlations (Full Correlation Matrix, bottom left, Reduced Correlation Matrix, bottom right).

## 4.4 Sioux-Falls network

The last test has been run on the Sioux-Falls network (South Dakota, US) (FIGURE 10), with reference to a single o-d pair **1-15**, characterized by 16 efficient routes, i.e. the o-d with the biggest choice set cardinality.

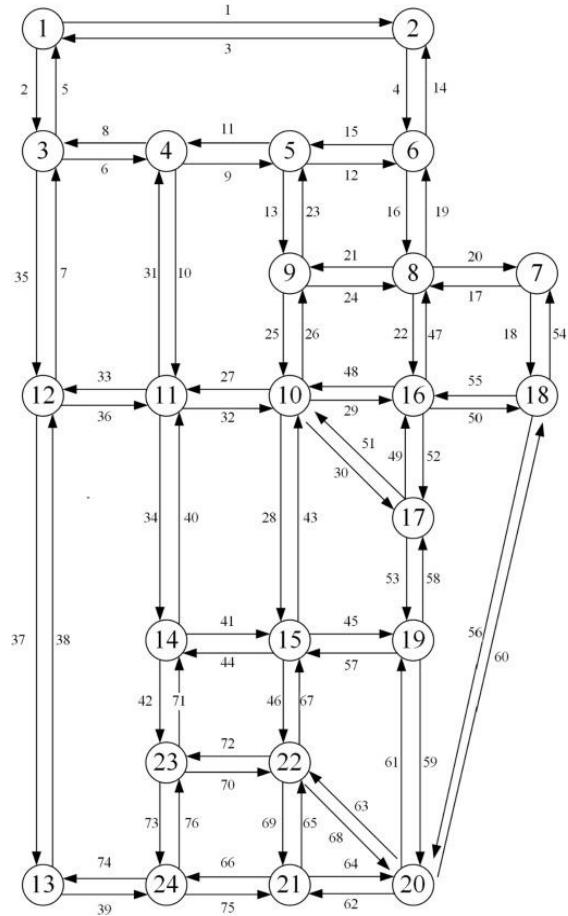

**FIGURE 10:** Sioux-Falls network, South Dakota (USA).

Apart from providing further evidence of the already observed performance of tested models, this network magnifies the effectiveness of the newly proposed specifications of $f(\mathbf{c}^i)$ in the weights of the CoNL, that exhibit significant improvements in fitting both MNP correlations and MNP probabilities (see FIGURE 11 and TABLE 8 in the Appendix). Furthermore, the LNL attains best fitting amongst testes models again with $\delta_{min}=0.4$, slightly outperforming the CoNL that yields best fitting for $0.2 \leq \delta_{min} \leq 0.3$ ($cv=0.1$) and $0.3 \leq \delta_{min} \leq 0.4$ ($cv=0.2$). However, CoNL results are more stable with respect to LNL when $\delta_{min}$ is varied.

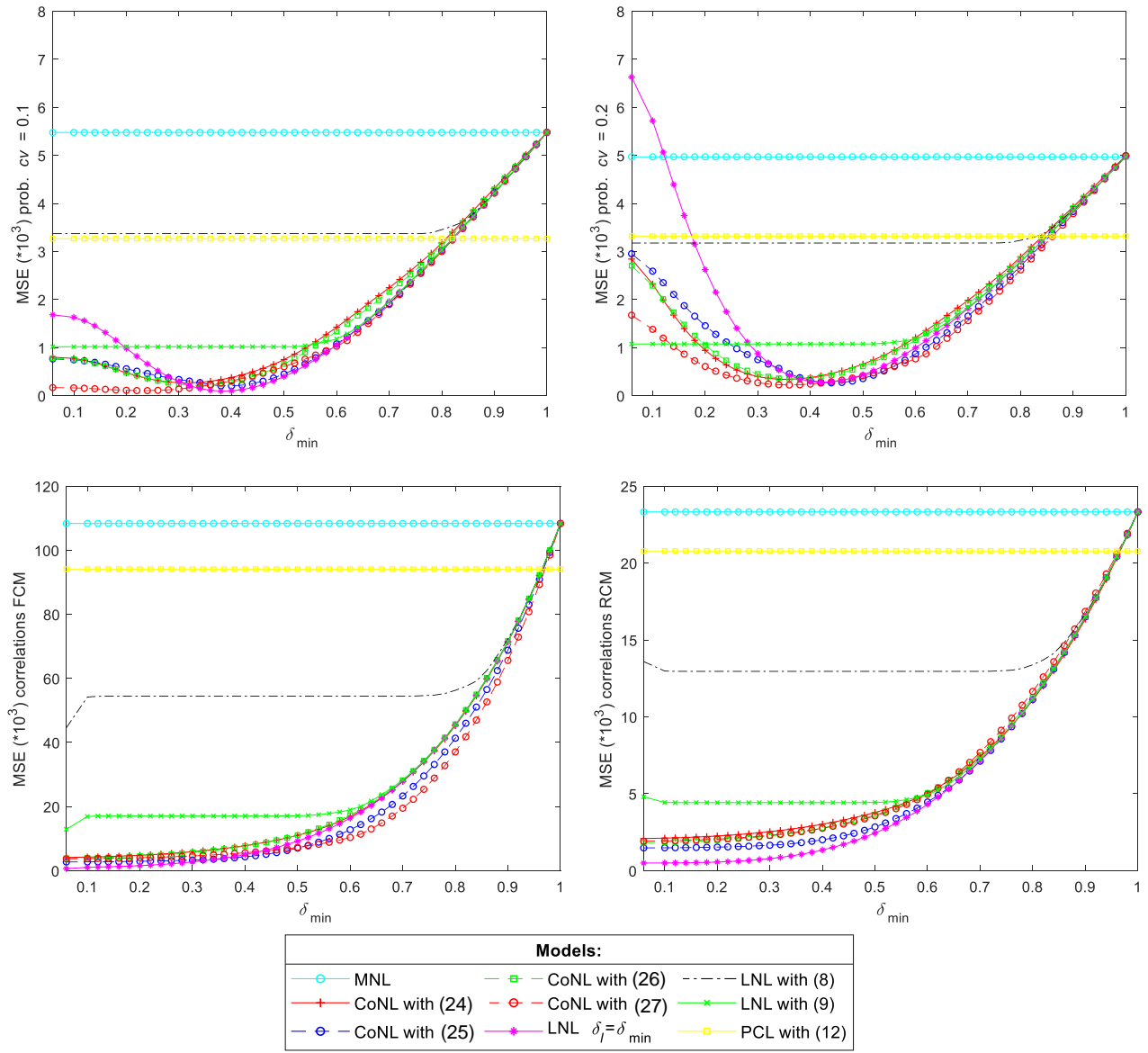

**FIGURE 11:** Sioux-Falls network, o-d pair *1-15* – Distance between tested models and MNP in terms of probabilities (cv=0.1 top left, cv=0.2 top right) and correlations (Full Correlation Matrix, bottom left, Reduced Correlation Matrix, bottom right).

# 5. Conclusions

This paper has investigated the performance of existing and new specifications of closed-form route choice models with flexible correlation patterns, namely the Link Nested Logit (LNL), the Paired Combinatorial Logit (PCL) and the more recent Combination of Nested Logit (CoNL) models. A well-consolidated research track on route choice modelling assumes that the topological overlapping of routes, often modelled through assumptions on the correlation structure of route choice models, has a remarkable impact on choices. As a consequence, a route choice model should exhibit sufficient flexibility in its underlying correlation pattern. In this respect, again following a consolidated assumption in the literature, choice probabilities and correlations of the Multinomial Probit (MNP) model by Daganzo & Sheffi (1977) are taken as target. Laboratory experiments on small/medium-size networks are illustrated, also leveraging a procedure for practical calculation of correlations of any GEV models, proposed by Marzano (2014).

This allowed answering small yet distinct research questions, that is:
- evaluating actual correlations underlying GEV route choice models with full correlation patterns;
- including the CoNL model in a comparison that includes GEV and MNP route choice models, never proposed in previous studies, except for Papola et al. (2018) only in terms of choice probabilities;
- testing novel and more effective specifications of the structural parameters of GEV and CoNL models, including fixing a lower bound $\delta_{min}$ for the nesting parameters of LNL and CoNL models.

Thanks to the set of the above-mentioned laboratory experiments on small/medium size networks, some distinct findings have been achieved. First, a model capable to account for general correlation patterns should be applied in any case to route choice modelling: this is clearly evident in all figures of the paper reporting, for each tested network, the trends of the distances between correlations/probabilities of MNP and correlations/probabilities of all the other models. Indeed, the PCL model fails in targeting MNP correlations, because known to exhibit a limited coverage of all feasible correlations, and consistently also always fails to target MNP probabilities. Importantly, for the LNL and CoNL models, a proper setting of a lower bound $\delta_{min}$ to the nesting parameters is crucial to achieve effective model performance: in particular, $\delta_{min}$ should be not too close to zero, otherwise LNL and CoNL models would become unstable, and not too great, because it would shrink too much the domain of feasible correlations allowed by the models. The experiments presented in the paper yielded an optimal setting of $\delta_{min}$ in between 0.2 and 0.4, always a little bit larger for the LNL than for the CoNL. Furthermore, outcomes of the experiments presented in the paper show that the CoNL performance is generally more stable than those of the other tested route choice models.

The CoNL benefits from the new proposed specifications of its weights, leading to results that improve those achieved by Papola et al. (2018). Overall, the CoNL is the best model in fitting MNP probabilities and correlations on the proposed test site. Notably, an LNL model specification with cross-nested structure as in Vovsha & Bekhor (1998) and nesting parameters equal to $\delta_{min}$ for all nests also provides satisfactory fitting of MNP probabilities and correlations.

Finally, the paper suggests a straightforward research prospect, which is replicating the study by estimating the same tested route choice models on a real-world dataset of trajectories (i.e. observed route choices) to further investigate their goodness-of-fit.

# Appendix: MSE tables

**TABLE 5**: Regular mesh network, o-d pair *1-9,* case #1– MSE values for the tested route choice models.

| | MNL | | | | CoNL (any) | | | |
|---|---|---|---|---|---|---|---|---|
| | Correlations | | Choice prob. | | Correlations | | Choice prob. | |
| | FCM | RCM | $cv=0.1$ | $cv=0.2$ | FCM | RCM | $cv=0.1$ | $cv=0.2$ |
| $\delta_{min}$ | MSE ($*10^3$) | | MSE ($*10^4$) | | MSE ($*10^3$) | | MSE ($*10^4$) | |
| 0 | | | | | **0.00** | **0.00** | 0.47 | 0.48 |
| 0.1 | | | | | 0.01 | 0.01 | 0.29 | 0.30 |
| 0.2 | | | | | 0.16 | 0.23 | **0.13** | **0.13** |
| 0.3 | | | | | 0.79 | 1.05 | 0.16 | 0.16 |
| 0.4 | | | | | 2.49 | 2.96 | 0.38 | 0.36 |
| 0.5 | 97.22 | 45.44 | 5.11 | 5.06 | 6.08 | 6.33 | 0.77 | 0.75 |
| 0.6 | | | | | 12.60 | 11.30 | 1.33 | 1.30 |
| 0.7 | | | | | 23.34 | 17.87 | 2.05 | 2.02 |
| 0.8 | | | | | 39.82 | 25.86 | 2.93 | 2.89 |
| 0.9 | | | | | 63.79 | 35.11 | 3.95 | 3.91 |
| 1 | | | | | 97.22 | 45.44 | 5.11 | 5.06 |

| | LNL ($\delta_l=\delta_{min}$) | | | | LNL with (8) | | | | LNL with (9) | | | | PCL with (12) | | | |
|---|---|---|---|---|---|---|---|---|---|---|---|---|---|---|---|---|
| | Correlations | | Choice prob. | | Correlations | | Choice prob. | | Correlations | | Choice prob. | | Correlations | | Choice prob. | |
| | FCM | RCM | $cv=0.1$ | $cv=0.2$ | FCM | RCM | $cv=0.1$ | $cv=0.2$ | FCM | RCM | $cv=0.1$ | $cv=0.2$ | FCM | RCM | $cv=0.1$ | $cv=0.2$ |
| $\delta_{min}$ | MSE ($*10^3$) | | MSE ($*10^4$) | | MSE ($*10^3$) | | MSE ($*10^4$) | | MSE ($*10^3$) | | MSE ($*10^4$) | | MSE ($*10^3$) | | MSE ($*10^4$) | |
| 0 | **0.92** | **1.00** | 2.35 | 2.39 | | | | | | | | | | | | |
| 0.1 | 1.03 | 1.13 | 1.56 | 1.59 | | | | | | | | | | | | |
| 0.2 | 1.51 | 1.67 | 0.64 | 0.66 | | | | | 8.44 | 8.04 | 0.44 | 0.43 | | | | |
| 0.3 | 2.57 | 2.80 | 0.19 | 0.20 | 32.24 | 22.32 | 2.30 | 2.27 | | | | | | | | |
| 0.4 | 4.66 | 4.83 | **0.15** | **0.14** | | | | | | | | | | | | |
| 0.5 | 8.44 | 8.04 | 0.44 | 0.43 | | | | | | | | | 60.86 | 34.05 | 1.38 | 2.75 |
| 0.6 | 14.85 | 12.63 | 1.02 | 1.00 | | | | | 14.85 | 12.64 | 1.02 | 1.00 | | | | |
| 0.7 | 25.15 | 18.72 | 1.83 | 1.80 | | | | | 25.15 | 18.72 | 1.83 | 1.80 | | | | |
| 0.8 | 40.92 | 26.28 | 2.81 | 2.78 | 40.92 | 26.28 | 2.81 | 2.78 | 40.92 | 26.28 | 2.81 | 2.78 | | | | |
| 0.9 | 64.15 | 35.22 | 3.92 | 3.87 | 64.15 | 35.22 | 3.92 | 3.87 | 64.15 | 35.22 | 3.92 | 3.87 | | | | |
| 1 | 97.22 | 45.44 | 5.11 | 5.06 | 97.22 | 45.44 | 5.11 | 5.06 | 97.22 | 45.44 | 5.11 | 5.06 | | | | |

**TABLE 6:** Regular mesh network, o-d pair *1-9,* case #2– MSE values for the tested route choice models.

| | MNL | | | | CoNL (any) | | | |
|---|---|---|---|---|---|---|---|---|
| | Correlations | | Choice prob. | | Correlations | | Choice prob. | |
| | FCM | RCM | $cv$=0.1 | $cv$=0.2 | FCM | RCM | $cv$=0.1 | $cv$=0.2 |
| $\delta_{min}$ | MSE (*$10^3$) | | MSE (*$10^4$) | | MSE (*$10^3$) | | MSE (*$10^4$) | |
| 0 | | | | | **1.68** | 3.35 | 5.64 | 9.74 |
| 0.1 | | | | | 1.80 | 3.29 | 5.50 | 8.25 |
| 0.2 | | | | | 2.28 | **3.20** | 3.75 | 5.04 |
| 0.3 | | | | | 3.30 | 3.45 | 1.96 | 2.80 |
| 0.4 | | | | | 5.20 | 4.41 | 1.06 | 1.41 |
| 0.5 | 86.96 | 36.63 | 6.12 | 8.78 | 8.57 | 6.49 | **0.76** | **0.77** |
| 0.6 | | | | | 14.12 | 9.80 | 0.83 | 0.80 |
| 0.7 | | | | | 23.25 | 14.62 | 1.33 | 1.62 |
| 0.8 | | | | | 37.60 | 20.93 | 2.48 | 3.37 |
| 0.9 | | | | | 58.28 | 28.32 | 4.09 | 5.79 |
| 1 | | | | | 86.96 | 36.63 | 6.12 | 8.78 |

| | LNL ($\delta_l=\delta_{min}$) | | | | LNL with (8) | | | | LNL with (9) | | | | PCL with (12) | | | |
|---|---|---|---|---|---|---|---|---|---|---|---|---|---|---|---|---|
| | Correlations | | Choice prob. | | Correlations | | Choice prob. | | Correlations | | Choice prob. | | Correlations | | Choice prob. | |
| | FCM | RCM | $cv$=0.1 | $cv$=0.2 | FCM | RCM | $cv$=0.1 | $cv$=0.2 | FCM | RCM | $cv$=0.1 | $cv$=0.2 | FCM | RCM | $cv$=0.1 | $cv$=0.2 |
| $\delta_{min}$ | MSE (*$10^3$) | | MSE (*$10^4$) | | MSE (*$10^3$) | | MSE (*$10^4$) | | MSE (*$10^3$) | | MSE (*$10^4$) | | MSE (*$10^3$) | | MSE (*$10^4$) | |
| 0 | **0.70** | **3.18** | 9.96 | 11.82 | **25.39** | 16.80 | | | **6.79** | 6.73 | | | | | | |
| 0.1 | 0.96 | 3.18 | 9.77 | 10.08 | | | | | | | | 1.07 | | | | |
| 0.2 | 1.40 | 3.28 | 6.77 | 5.19 | | | | | 7.41 | **6.65** | 0.86 | | | | | |
| 0.3 | 2.36 | 3.71 | 3.32 | 2.60 | 26.64 | **16.61** | 2.18 | 2.33 | | | | | | | | |
| 0.4 | 4.23 | 4.84 | 1.48 | 1.15 | | | | | 7.53 | 6.74 | | 0.89 | | | | |
| 0.5 | 7.62 | 7.00 | **0.86** | **0.58** | | | | | 9.04 | 7.84 | | 0.59 | 54.58 | 27.52 | 4.42 | 8.82 |
| 0.6 | 13.35 | 10.42 | 0.98 | 0.83 | 26.68 | 16.63 | | | 13.78 | 10.69 | 1.06 | 0.93 | | | | |
| 0.7 | 22.55 | 15.18 | 1.63 | 1.83 | 28.74 | 17.77 | 2.38 | 2.71 | 22.55 | 15.18 | 1.63 | 1.83 | | | | |
| 0.8 | 36.65 | 21.23 | 2.72 | 3.54 | 38.64 | 22.02 | 3.04 | 3.88 | 36.65 | 21.23 | 2.72 | 3.54 | | | | |
| 0.9 | 57.40 | 28.43 | 4.22 | 5.88 | 57.40 | 28.43 | 4.22 | 5.88 | 57.40 | 28.43 | 4.22 | 5.88 | | | | |
| 1 | 86.96 | 36.63 | 6.12 | 8.78 | 86.96 | 36.63 | 6.12 | 8.78 | 86.96 | 36.63 | 6.12 | 8.78 | | | | |

**TABLE 7:** Mesh network with bypass, o-d pair *1-12* – MSE values for the tested route choice models.

|  | MNL | | | | CoNL with (24) | | | | CoNL with (25) | | | | CoNL with (26) | | | | CoNL with (27) | | | |
|---|---|---|---|---|---|---|---|---|---|---|---|---|---|---|---|---|---|---|---|---|
|  | Correlations | | Choice prob. | | Correlations | | Choice prob. | | Correlations | | Choice prob. | | Correlations | | Choice prob. | | Correlations | | Choice prob. | |
|  | FCM | RCM | $cv=0.1$ | $cv=0.2$ | FCM | RCM | $cv=0.1$ | $cv=0.2$ | FCM | RCM | $cv=0.1$ | $cv=0.2$ | FCM | RCM | $cv=0.1$ | $cv=0.2$ | FCM | RCM | $cv=0.1$ | $cv=0.2$ |
| $\delta_{min}$ | MSE ($*10^3$) | | MSE ($*10^4$) | | MSE ($*10^3$) | | MSE ($*10^4$) | | MSE ($*10^3$) | | MSE ($*10^4$) | | MSE ($*10^3$) | | MSE ($*10^4$) | | MSE ($*10^3$) | | MSE ($*10^4$) | |
| 0 |  |  |  |  | **3.73** | **1.47** | 0.97 | 1.56 | **1.53** | **0.51** | 1.81 | 5.54 | **2.70** | **0.75** | 1.47 | 5.04 | **1.52** | **0.58** | 1.58 | 5.11 |
| 0.1 |  |  |  |  | 3.87 | 1.51 | 0.88 | 1.41 | 1.64 | 0.59 | 1.65 | 4.38 | 2.88 | 0.85 | 1.28 | 3.80 | 1.61 | 0.66 | 1.43 | 3.94 |
| 0.2 |  |  |  |  | 4.40 | 1.68 | 0.72 | 0.94 | 2.12 | 0.93 | 0.91 | 1.33 | 3.59 | 1.26 | 0.66 | 1.03 | 2.02 | 1.01 | 0.77 | 1.08 |
| 0.3 |  |  |  |  | 5.65 | 2.37 | **0.25** | **0.24** | 3.31 | 1.72 | **0.22** | **0.23** | 5.09 | 2.14 | **0.21** | **0.22** | 2.93 | 1.72 | **0.17** | **0.15** |
| 0.4 |  |  |  |  | 8.72 | 3.81 | 0.45 | 0.30 | 6.20 | 3.30 | 0.36 | **0.23** | 8.06 | 3.72 | 0.47 | 0.32 | 4.84 | 3.00 | 0.30 | 0.19 |
| 0.5 | **158.50** | **53.08** | **9.62** | **5.49** | 14.34 | 6.24 | 1.33 | 0.79 | 11.78 | 5.95 | 1.33 | 0.78 | 13.71 | 6.33 | 1.42 | 0.85 | 10.01 | 5.54 | 1.34 | 0.80 |
| 0.6 |  |  |  |  | 24.21 | 10.12 | 2.63 | 1.51 | 22.20 | 10.28 | 2.85 | 1.65 | 23.84 | 10.52 | 2.82 | 1.63 | 20.15 | 9.90 | 2.88 | 1.65 |
| 0.7 |  |  |  |  | 41.37 | 16.24 | 4.33 | 2.46 | 39.67 | 16.77 | 4.56 | 2.58 | 41.31 | 16.95 | 4.52 | 2.56 | 37.43 | 16.40 | 4.58 | 2.59 |
| 0.8 |  |  |  |  | 67.66 | 25.04 | 6.13 | 3.45 | 66.32 | 25.75 | 6.29 | 3.55 | 67.79 | 25.86 | 6.26 | 3.53 | 64.23 | 25.44 | 6.31 | 3.55 |
| 0.9 |  |  |  |  | 105.68 | 37.11 | 7.91 | 4.47 | 104.88 | 37.69 | 7.99 | 4.52 | 105.84 | 37.74 | 7.98 | 4.51 | 103.47 | 37.50 | 8.00 | 4.53 |
| 1 |  |  |  |  | 158.50 | 53.08 | 9.62 | 5.49 | 158.50 | 53.08 | 9.62 | 5.49 | 158.50 | 53.08 | 9.62 | 5.49 | 158.50 | 53.08 | 9.62 | 5.49 |

|  | LNL ($\delta_l=\delta_{min}$) | | | | LNL with (8) | | | | LNL with (9) | | | | PCL with (12) | | | |
|---|---|---|---|---|---|---|---|---|---|---|---|---|---|---|---|---|
|  | Correlations | | Choice prob. | | Correlations | | Choice prob. | | Correlations | | Choice prob. | | Correlations | | Choice prob. | |
|  | FCM | RCM | $cv=0.1$ | $cv=0.2$ | FCM | RCM | $cv=0.1$ | $cv=0.2$ | FCM | RCM | $cv=0.1$ | $cv=0.2$ | FCM | RCM | $cv=0.1$ | $cv=0.2$ |
| $\delta_{min}$ | MSE ($*10^3$) | | MSE ($*10^4$) | | MSE ($*10^3$) | | MSE ($*10^4$) | | MSE ($*10^3$) | | MSE ($*10^4$) | | MSE ($*10^3$) | | MSE ($*10^4$) | |
| 0 | **0.65** | **0.58** | 4.79 | 14.81 | **61.18** | **18.48** | **5.37** | 4.18 | **17.15** | **5.83** | 1.46 | 2.00 |  |  |  |  |
| 0.1 | 1.07 | 0.65 | 4.36 | 12.47 | 64.45 | 18.08 | **5.37** | 4.18 | 18.69 | 5.71 | 1.46 | 1.90 |  |  |  |  |
| 0.2 | 1.71 | 1.08 | 2.42 | 4.08 | 64.48 | 18.21 | 5.44 | 4.13 | 18.71 | 5.97 | **1.44** | 1.41 |  |  |  |  |
| 0.3 | 3.27 | 2.01 | 0.47 | 0.65 | 64.52 | 18.54 | 5.61 | 4.09 | 18.74 | 6.44 | 1.46 | 1.19 |  |  |  |  |
| 0.4 | 6.52 | 3.69 | **0.10** | **0.09** | 64.61 | 19.11 | 5.78 | **4.01** | 18.83 | 7.16 | 1.56 | **1.10** |  |  |  |  |
| 0.5 | 12.60 | 6.43 | 0.98 | 0.55 | 64.79 | 19.97 | 5.91 | 3.89 | 19.96 | 8.44 | 1.82 | 1.13 | 140.38 | 48.00 | 5.72 | 2.55 |
| 0.6 | 23.12 | 10.65 | 2.49 | 1.38 | 65.10 | 21.18 | 5.98 | 3.73 | 26.55 | 11.46 | 2.75 | 1.55 |  |  |  |  |
| 0.7 | 40.12 | 16.83 | 4.25 | 2.36 | 65.81 | 22.84 | 6.02 | 3.57 | 41.34 | 17.03 | 4.26 | 2.36 |  |  |  |  |
| 0.8 | 66.18 | 25.53 | 6.08 | 3.40 | 74.74 | 27.01 | 6.47 | 3.67 | 66.18 | 25.53 | 6.08 | 3.40 |  |  |  |  |
| 0.9 | 104.40 | 37.39 | 7.89 | 4.45 | 106.11 | 37.46 | 7.86 | 4.42 | 104.40 | 37.39 | 7.89 | 4.45 |  |  |  |  |
| 1 | 158.50 | 53.08 | 9.62 | 5.49 | 158.50 | 53.08 | 9.62 | 5.49 | 158.50 | 53.08 | 9.62 | 5.49 |  |  |  |  |

**TABLE 8:** Sioux-Falls network, o-d pair *1-15* – MSE values for the tested route choice models.

| | MNL | | | | CoNL with (24) | | | | CoNL with (25) | | | | CoNL with (26) | | | | CoNL with (27) | | | |
|---|---|---|---|---|---|---|---|---|---|---|---|---|---|---|---|---|---|---|---|---|
| | **Correlations** | | **Choice prob.** | | **Correlations** | | **Choice prob.** | | **Correlations** | | **Choice prob.** | | **Correlations** | | **Choice prob.** | | **Correlations** | | **Choice prob.** | |
| | **FCM** | **RCM** | $cv=0.1$ | $cv=0.2$ | **FCM** | **RCM** | $cv=0.1$ | $cv=0.2$ | **FCM** | **RCM** | $cv=0.1$ | $cv=0.2$ | **FCM** | **RCM** | $cv=0.1$ | $cv=0.2$ | **FCM** | **RCM** | $cv=0.1$ | $cv=0.2$ |
| $\delta_{min}$ | MSE ($*10^3$) | | MSE ($*10^4$) | | MSE ($*10^3$) | | MSE ($*10^4$) | | MSE ($*10^3$) | | MSE ($*10^4$) | | MSE ($*10^3$) | | MSE ($*10^4$) | | MSE ($*10^3$) | | MSE ($*10^4$) | |
| 0 | | | | | **4.12** | **2.09** | 0.80 | 2.84 | **2.70** | **1.48** | 0.77 | 2.95 | **3.87** | **1.81** | 0.78 | 2.70 | **3.69** | **1.91** | 0.17 | 1.68 |
| 0.1 | | | | | 4.23 | 2.11 | 0.78 | 2.32 | 2.73 | 1.49 | 0.75 | 2.59 | 3.98 | 1.84 | 0.76 | 2.29 | 3.75 | 1.94 | 0.16 | 1.38 |
| 0.2 | | | | | 4.80 | 2.25 | 0.48 | 0.94 | 2.91 | 1.53 | 0.56 | 1.46 | 4.55 | 1.98 | 0.48 | 1.05 | 4.04 | 2.05 | **0.11** | 0.61 |
| 0.3 | | | | | 5.91 | 2.53 | **0.27** | 0.40 | 3.30 | 1.65 | 0.34 | 0.75 | 5.67 | 2.25 | **0.24** | 0.45 | 4.60 | 2.29 | 0.14 | 0.27 |
| 0.4 | | | | | 7.80 | 3.02 | 0.38 | **0.38** | 4.31 | 2.01 | **0.20** | **0.32** | 7.67 | 2.78 | 0.28 | **0.36** | 5.58 | 2.75 | 0.31 | **0.24** |
| 0.5 | **108.35** | **23.33** | 5.48 | 4.97 | 11.02 | 3.78 | 0.75 | 0.66 | 7.03 | 2.84 | 0.45 | 0.36 | 11.08 | 3.62 | 0.66 | 0.61 | 7.25 | 3.56 | 0.60 | 0.42 |
| 0.6 | | | | | 16.89 | 5.04 | 1.42 | 1.21 | 12.76 | 4.46 | 1.08 | 0.87 | 17.36 | 5.05 | 1.34 | 1.16 | 10.35 | 4.96 | 1.03 | 0.77 |
| 0.7 | | | | | 27.86 | 7.39 | 2.25 | 1.98 | 23.31 | 7.12 | 1.92 | 1.66 | 28.24 | 7.51 | 2.16 | 1.91 | 19.49 | 7.69 | 1.90 | 1.56 |
| 0.8 | | | | | 45.24 | 11.09 | 3.18 | 2.89 | 41.34 | 11.13 | 3.03 | 2.70 | 45.71 | 11.30 | 3.14 | 2.85 | 37.06 | 11.65 | 3.00 | 2.62 |
| 0.9 | | | | | 71.38 | 16.36 | 4.31 | 3.93 | 68.81 | 16.50 | 4.23 | 3.83 | 71.74 | 16.54 | 4.29 | 3.91 | 65.60 | 16.86 | 4.21 | 3.79 |
| 1 | | | | | 108.35 | 23.33 | 5.48 | 4.99 | 108.35 | 23.33 | 5.48 | 4.99 | 108.35 | 23.33 | 5.48 | 4.98 | 108.35 | 23.33 | 5.48 | 5.00 |

| | LNL ($\delta_l=\delta_{min}$) | | | | LNL with (8) | | | | LNL with (9) | | | | PCL with (12) | | | |
|---|---|---|---|---|---|---|---|---|---|---|---|---|---|---|---|---|
| | **Correlations** | | **Choice prob.** | | **Correlations** | | **Choice prob.** | | **Correlations** | | **Choice prob.** | | **Correlations** | | **Choice prob.** | |
| | **FCM** | **RCM** | $cv=0.1$ | $cv=0.2$ | **FCM** | **RCM** | $cv=0.1$ | $cv=0.2$ | **FCM** | **RCM** | $cv=0.1$ | $cv=0.2$ | **FCM** | **RCM** | $cv=0.1$ | $cv=0.2$ |
| $\delta_{min}$ | MSE ($*10^3$) | | MSE ($*10^4$) | | MSE ($*10^3$) | | MSE ($*10^4$) | | MSE ($*10^3$) | | MSE ($*10^4$) | | MSE ($*10^3$) | | MSE ($*10^4$) | |
| 0 | **0.65** | **0.50** | 1.68 | 6.63 | **44.59** | 13.60 | | | **12.89** | 4.80 | | | | | | |
| 0.1 | 1.02 | 0.50 | 1.63 | 5.72 | 54.18 | 12.98 | | | 16.93 | 4.43 | | | | | | |
| 0.2 | 1.56 | 0.56 | 0.99 | 2.62 | | | | | 17.06 | | 1.02 | 1.08 | | | | |
| 0.3 | 2.72 | 0.78 | 0.29 | 0.88 | | | 3.37 | 3.18 | 17.06 | **4.42** | | | | | | |
| 0.4 | 5.02 | 1.34 | **0.10** | **0.30** | 54.45 | **12.96** | | | 17.06 | | | | **94.11** | **20.78** | **3.27** | **3.32** |
| 0.5 | 9.22 | 2.44 | 0.40 | 0.44 | | | | | 17.14 | 4.44 | | | | | | |
| 0.6 | 16.39 | 4.31 | 1.05 | 1.00 | | | | | 18.94 | 4.93 | 1.18 | 1.18 | | | | |
| 0.7 | 27.90 | 7.14 | 1.96 | 1.82 | | | | | 28.35 | 7.25 | 1.97 | 1.84 | | | | |
| 0.8 | 45.54 | 11.15 | 3.03 | 2.79 | 56.28 | 13.38 | 3.45 | 3.23 | 45.55 | 11.15 | 3.03 | 2.79 | | | | |
| 0.9 | 71.49 | 16.50 | 4.22 | 3.85 | 72.29 | 16.69 | 4.23 | 3.87 | 71.49 | 16.50 | 4.22 | 3.85 | | | | |
| 1 | 108.35 | 23.33 | 5.48 | 4.97 | 108.35 | 23.33 | 5.48 | 4.97 | 108.35 | 23.33 | 5.48 | 4.97 | | | | |

# CRediT Author Statement

**Fiore Tinessa**: Conceptualization, Data curation, Formal analysis, Methodology, Software, Writing (original draft), Writing (review and editing). **Vittorio Marzano**: Conceptualization, Methodology, Software, Writing (original draft), Writing (review and editing). **Andrea Papola**: Conceptualization, Methodology, Writing (review and editing), Funding acquisition, Supervision.

Chu, C. (1989). A paired combinatorial logit model for travel demand analysis. *5th World Conference on Transportation Research*. Ventura, CA.

Daganzo, C. F., & Sheffi, Y. (1977). On stochastic models of traffic assignment. *Transportation Science*. https://doi.org/10.1287/trsc.11.3.253

Daly, A. (2001). *Recursive nested EV model*.

Daly, Andrew, & Bierlaire, M. (2006). A general and operational representation of Generalised Extreme Value models. *Transportation Research Part B: Methodological*, *40*(4), 285–305. https://doi.org/10.1016/j.trb.2005.03.003

Daly, Andrew, & Zachary, S. (1978). Improved multiple choice models. In *Determinants of travel choice*. Saxon House, Sussex.

Dial, R. B. (1971). A probabilistic multipath traffic assignment model which obviates path enumeration. *Transportation Research*. https://doi.org/10.1016/0041-1647(71)90012-8

Ding, C., Wang, Y., Yang, J., Liu, C., & Lin, Y. (2016). Spatial heterogeneous impact of built environment on household auto ownership levels: Evidence from analysis at traffic analysis zone scales. *Transportation Letters*. https://doi.org/10.1179/1942787515Y.0000000004

Fosgerau, M., & Bierlaire, M. (2009). Discrete choice models with multiplicative error terms. *Transportation Research Part B: Methodological*, *43*(5), 494–505. https://doi.org/10.1016/j.trb.2008.10.004

Fosgerau, Mogens, Frejinger, E., & Karlstrom, A. (2013). A link based network route choice model with unrestricted choice set. *Transportation Research Part B: Methodological*, *56*, 70–80. https://doi.org/10.1016/j.trb.2013.07.012

Frejinger, E., & Bierlaire, M. (2007). Capturing correlation with subnetworks in route choice models. *Transportation Research Part B: Methodological*, *41*(3), 363–378. https://doi.org/10.1016/j.trb.2006.06.003

Frejinger, E., Bierlaire, M., & Ben-Akiva, M. (2009). Sampling of alternatives for route choice modeling. *Transportation Research Part B: Methodological*, *43*(10), 984–994. https://doi.org/10.1016/j.trb.2009.03.001

Geweke, J. (1991). Efficient Simulation from the Multivariate Normal and Student-t Distributions Subject to Linear Constraints and the Evaluation of Constraint Probabilities. *Computing Science and Statistics: The 23rd Symposium on the Interface*. https://doi.org/10.1.1.26.6892

Gliebe, J. P., Koppelman, F. S., & Ziliaskopoulos, A. (1999). Route choice using a paired combinatorial logit model. *78th Meeting of the Transportation Research Board*. Washington D.C.

Haghani, M., Shahhoseini, Z., & Sarvi, M. (2016). Quantifying benefits of traveler information systems to performance of transport networks prior to implementation: A double-class structured-parameter stochastic trip assignment approach. *Transportation Letters*. https://doi.org/10.1179/1942787515Y.0000000013

Hajivassiliou, V., McFadden, D., & Ruud, P. (1996). Simulation of multivariate normal rectangle probabilities and their derivatives theoretical and computational results. *Journal of Econometrics*, *72*(1–2), 85–134. https://doi.org/10.1016/0304-4076(94)01716-6

Hood, J., Sall, E., & Charlton, B. (2011). A GPS-based bicycle route choice model for San Francisco, California. *Transportation Letters*. https://doi.org/10.3328/TL.2011.03.01.63-75

Hoogendoorn-Lanser, S. (2005). Modelling Travel Behaviour in Multi-modal Networks. *TRAIL Thesis Series*.

Horowitz, J. L. (1991). Reconsidering the multinomial probit model. *Transportation Research Part B*. https://doi.org/10.1016/0191-2615(91)90036-I

Horowitz, J. L., Sparmann, J. M., & Daganzo, C. F. (1982). INVESTIGATION OF THE ACCURACY OF THE CLARK APPROXIMATION FOR THE MULTINOMIAL PROBIT MODEL. *Transportation Science*. https://doi.org/10.1287/trsc.16.3.382

Huang, C., Burris, M., & Shaw, W. D. (2017). Models of transportation choice with risk: an application to managed lanes. *Transportation Letters*. https://doi.org/10.1080/19427867.2016.1204811

Kahneman, D., & Tversky, A. (1979). An analysis of decision under risk. *Econometrica*.

Kazagli, E., Bierlaire, M., & de Lapparent, M. (2020). Operational route choice methodologies for practical applications. *Transportation*, *47*(1), 43–74. https://doi.org/10.1007/s11116-017-9849-0

Kazagli, E., Bierlaire, M., & Flötteröd, G. (2016). Revisiting the route choice problem: A modeling framework based on mental representations. *Journal of Choice Modelling*, *19*, 1–23. https://doi.org/10.1016/j.jocm.2016.06.00129